\documentclass[aps,prb,twocolumn,groupedaddress,showpacs,superscriptaddress,amssymb,amsmath,longbibliography,]{revtex4-1}

\usepackage{physics}

\usepackage{amsmath}
\usepackage{mathptmx,bm}
\usepackage{graphicx}
\usepackage{dcolumn}
\usepackage{tabularx}
\usepackage{epsf}
\usepackage{color}

\usepackage{hyperref}
\hypersetup{breaklinks,colorlinks,linkcolor=blue,citecolor=blue,urlcolor=blue}

\usepackage{epstopdf}%
\usepackage{verbatim}
\usepackage[english]{babel}
\newcommand{\bea}{\begin{eqnarray}}
\newcommand{\eea}{\end{eqnarray}}
\usepackage{amsfonts}
\DeclareUnicodeCharacter{2009}{\,} 
\begin{document}

%
\title{Effective Hamiltonian theory of open quantum systems at strong coupling}

\author{Nicholas Anto-Sztrikacs}
\affiliation{Department of Physics, 60 Saint George St., University of Toronto, Toronto, Ontario, M5S 1A7, Canada}

\author{Ahsan Nazir}
\affiliation{Department of Physics and Astronomy, University of Manchester, Oxford Road, Manchester, M13 9PL, United Kingdom}

\author{Dvira Segal}
\affiliation{Chemical Physics Theory Group, Department of Chemistry and Centre for Quantum Information and Quantum Control,
University of Toronto, 80 Saint George St., Toronto, Ontario, M5S 3H6, Canada}
\affiliation{Department of Physics, 60 Saint George St., University of Toronto, Toronto, Ontario, M5S 1A7, Canada}
\email{dvira.segal@utoronto.ca}

\date{\today}

\begin{abstract}
We present the reaction-coordinate polaron-transform (RCPT) framework for generating effective Hamiltonian models to treat
nonequilibrium open quantum systems at strong coupling with their surroundings. 
Our approach, which is based on two exact transformations of the Hamiltonian followed by its controlled 
truncation, ends with a new Hamiltonian with a weakened coupling to the environment.
This new effective Hamiltonian mirrors the initial one,
 except that its parameters are dressed by the system-bath couplings. 
The power and elegance of the RCPT approach lie in its generality and in its mathematical simplicity, allowing for analytic work and thus profound understanding 
of the impact of strong system-bath coupling effects on open quantum system phenomena.
Examples interrogated in this work include canonical models for quantum thermalization,
 charge and energy transport at the nanoscale, performance bounds of
quantum thermodynamical machines such as absorption refrigerators and thermoelectric generators, as well as the equilibrium and nonequilibrium behavior of many-body dissipative spin chains.
%
%
\end{abstract}

\maketitle


\section{Introduction}
\label{Sec:Introduction}

Quantum systems are inevitably coupled to their surrounding environment. 
At the nanoscale, these interactions are influential and cannot be neglected, 
which in turn leads to theoretical and technical challenges in modelling open quantum systems. 
Quantum master equation (QME) approaches offer a powerful framework
for simulating open quantum systems. 
While the Nakajima-Zwanzig formalism is exact \cite{Nitzan}, approximations must be made for practical computations.
Most commonly, QMEs are made perturbative in the system-bath coupling parameter; 
the prominent Redfield equation takes into account only
the lowest (second) nontrivial order in this expansion, referred to as the Born approximation.
Weak coupling QMEs offer straightforward computations and analytical results in some cases, and 
as such they have gained enormous popularity in diverse fields: e.g. chemical dynamics \cite{Nitzan}, quantum optics \cite{Breuer}, 
quantum information science \cite{Lidar,Nielson}, 
and quantum thermodynamics \cite{Kosloff_2013,Kosloff_2018}.
However, these methods are strictly limited to the weak coupling regime, missing rich physics.
This work presents a Hamiltonian reformation (transformation and truncation) technique that allows 
treating strong coupling regimes 
while providing both 
detailed understanding of such effects in quantum systems, 
and a 
cheap route for computations.
%
Applications detailed in this work concern quantum transport and quantum thermodynamics problems 
\citep{Anders_2016,Goold_2016,Kosloff_2013,Kosloff_2018,Levy14,Luis14,Strasberg_2021,Deffner_2019}, where a consistent theory of 
thermodynamics in the quantum regime relies on the correct treatment of strong coupling features. 
However, our approach can be exercised on other open problems in a variety of contexts.

Focusing on quantum thermodynamics in the context of thermal machines, 
strong coupling effects can allow non-classical correlations to build-up between the system and its reservoirs, which
could be utilized as a resource to design novel quantum technologies  \citep{NazirPRA14,Rivas_2020}.  
While it is debated whether strong coupling effects are beneficial or detrimental to their performance \cite{JunjieA1,JunjieA2,Hava2018,Scarani2018},
it is clear that strong coupling effects can significantly impact the performance and efficiency of thermal devices
e.g., by renormalizing parameters and opening new transport pathways 
\citep{QAR-Felix,Newman_2017,Mu_2017}.  
Additionally, in the context of thermalization, strong system-bath interactions lead to deviations from the canonical distribution, predicted under the  assumption of vanishingly-weak system-reservoir coupling \citep{NazirPRA14,Giacomo,Archak,Cresser,Valkunas20,Zhang_2022,Trusheckin_2022,Kawai_2019,Kawai2020,Keeling2022,Cheng2022}. Strong coupling is also responsible for deviations from simple additivity approximations \citep{Ahsan2019,Ahsan2022,Gernot17}.

To capture such non-trivial effects, one must go beyond second-order perturbative QMEs. One such choice are numerically-exact methods, including the 
multiconfiguration time-dependent Hartree (MCTDH) approach \cite{MCTDH1,MCTDH2,MCTDH3}, numerical renormalization group methods \cite{NRG1,NRG2,NRG3},
the hierarchical equations of motion \citep{TanimuraBook,Tanimura20}, path integral approaches \citep{Leggett,Makri1,Makri2,Makri20a,Makri20b,Thorwart,SiminePI,Kilgour19,Keeling}, quantum Monte Carlo algorithms \cite{QMC1,QMC2,QMC3}, 
chain mapping techniques \citep{TEDOPA1,TEDOPA2,chainPlenio,chainDario}, tensor network based methods \citep{Brenes_2020}, and more. Though these methods provide accurate benchmarks for describing open quantum systems at strong coupling, (i) they  are often limited to minimal models and (ii) fail to provide analytical intuition and hence, do not allow us to pursue the objective of this work: To understand the fundamental essence of strong coupling. 
Conversely, there exist other inexact tools that allow the development of analytic understanding. This includes the  noninteracting blip approximation (NIBA), which is accurate for Ohmic spectral functions \citep{Weiss,NIBA11,NIBA14}, the polaron-transformed Redfield equation, which allows for more general spectral functions \citep{Silbey,Dave08,Cao12,Cao1,Cao2}, and Greens functions techniques \citep{NEGFBF,MW,Misha,ThossNEGF,WuNEGF,NJP17,PT-NEGF17}. However, these tools typically become immediately cumbersome beyond minimal models, and are restricted in their applicability since they are perturbative in some parameters.

The reaction coordinate (RC) mapping \citep{NazirPRA14,Nazir18,Strasberg_2016,Burghardt1,Burghardt2} bridges the gap between powerful numerical tools, and low-order perturbative analytical methods. 
While originally developed in the context of chemical reactivity \cite{Burghardt1,Burghardt2}, in recent years the method has found numerous applications 
in the context of quantum thermodynamics as a general tool to capture the effects of strong system-reservoir coupling.
In this technique, a central, collective degree of freedom from the reservoir is extracted and included as part of the system. The original quantum system is then extended, and termed an ``enlarged system". This enlarged system now contains the open system and the collective degree of freedom, the reaction coordinate, extracted from the bath. 
The RC mapping can be used in conjunction with weak coupling QME tools
since, with a proper choice of parameters, the coupling of the {\it enlarged} system  to the {\it residual} bath is weak.
The RC method treats {\it non-perturbatively} the  system-reservoir coupling parameter to obtain appropriate dynamical and steady state properties of the original open system. The combined RC-QME approach has been utilized for studying the quantum dynamics of impurity models \citep{NazirPRA14,Nazir16}, thermal transport in nanojunctions \citep{Nick2021,Nick2022,Correa19}, the operation of quantum thermal machines \citep{Newman_2020,McConnel22,QAR-Felix,Newman_2017}, transport in electronic systems \citep{GernotF,GernotF2,Ahsan2019B}, problems of equilibration \citep{NazirPRA14,Camille} as well as the dynamics of non-Markovian systems \citep{Nickdecoh}. As useful as this tool has been in recent years, and despite relying on an analytical mapping, this approach in practice has been only used as a numerical method  due to the large Hilbert space of the enlarged system.
  
The polaron or Lang-Firsov transformation is central to many-body physics, with applications extending far beyond the original coupled electron-phonon problem.
The concept of a polaron originates from solid state physics whereby lattice vibrations couple to the electron, generating a heavy electron with an effective mass representing the electron plus the phonon cloud surrounding it \citep{Mahan}. 
In the broader context of an impurity immersed in a harmonic bath, 
the polaron transformation allows to unitarily map the system to the polaron frame where the system Hamiltonian is dressed by the system-bath coupling, making it amenable to perturbative treatments in the so called nonadiabatic parameter (tunneling splitting). This approach has been used e.g., to simulate heat transport through quantum nanojunctions \cite{PRL05,QME06,NIBA11,NIBA14},
later further extended using the polaron-transformed QME
\citep{Cao1,Cao2,Cao3,Cao4}. However, despite providing analytical insight on spin-boson type models,
 the polaron mapping can become cumbersome with  compound (nonadditive) bath interaction terms. As such, it was mainly utilized in simplified models with independent baths locally affecting the system (see e.g., Refs. \citenum{SegalCP02,AhsanNJP13}), or by restricting the form of the system's Hamiltonian to eliminate the formation of composite interaction terms \cite{Renjie19}. 

\begin{figure*}
\centering
\includegraphics[width=2\columnwidth]{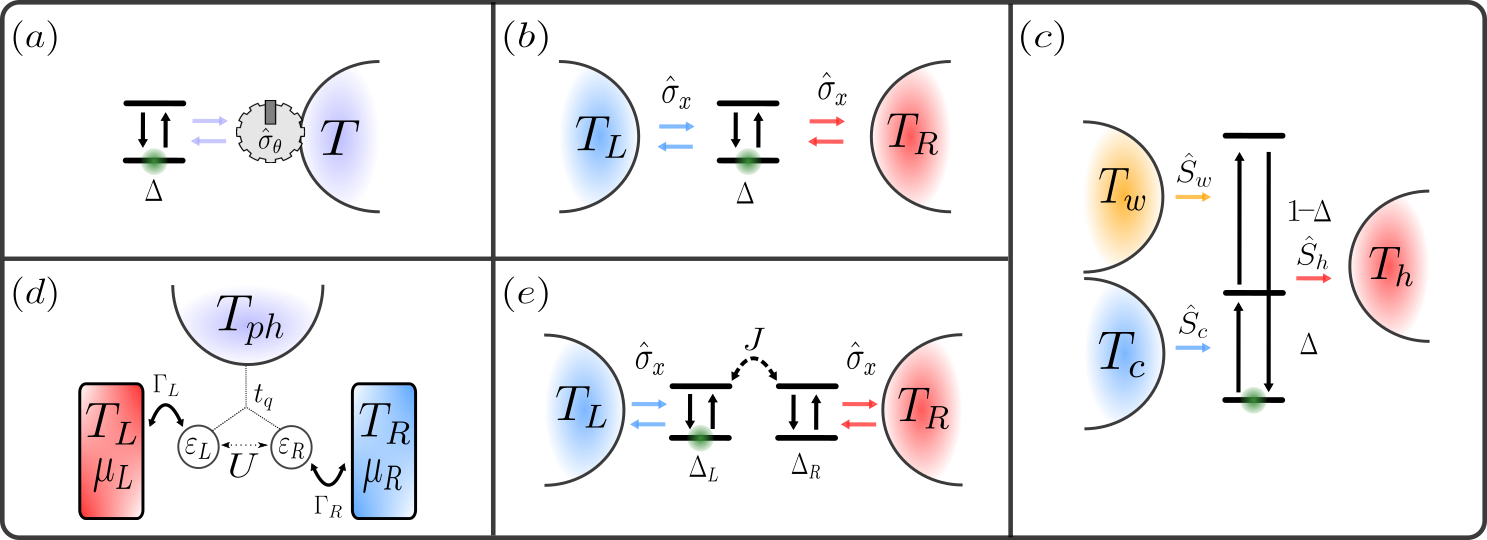}
\caption{Diagrams of the five problems examined in this work using the RCPT method.  
 (a) {\it Quantum thermalization}, studied with a two-level impurity spin coupled to a bosonic reservoir experiencing both decohering and energy-exchange effects. 
 (b) {\it Quantum heat transport}, examined on a minimal model of a spin system coupled to two bosonic reservoirs with a temperature bias. 
 (c) {\it Quantum refrigeration}, examined on a three-level quantum absorption refrigerator, where each transition couples to a different reservoir, resulting in the net effect of extracting heat from the cold environment. 
 (d) {\it Thermoelectric performance of nanojunctions}, illustrated 
 on a phonon-assisted double quantum dot nanojunction.
Here, charge is transported between the two leads with the help of a phonon environment. 
(e) {\it  Dissipative spin chains}, with each spin coupled to an independent heat bath, here illustrated with a two-qubit model.}
\label{fig:diagram-models}
\end{figure*}

In this work, we combine two central transformations in open quantum systems methodologies, namely, the reaction coordinate mapping and the polaron transformation, and develop a general and robust tool for {\it understanding} and {\it feasibly simulating} strong coupling features in quantum transport and thermodynamics. 
The essence of this newly-developed reaction coordinate polaron-transformation (RCPT) approach is 
that the succession of these two transformations {\it  imprints} strong coupling effects {\it directly} into the system's Hamiltonian, which after the transformations becomes {\it weakly} coupled to the (residual) surroundings, allowing the use of weak-coupling techniques. 
Furthermore, due to the additional truncation of the RC manifold, 
the dimensionality of the Hilbert space is identical to the starting one. 
Thus, the RCPT method allows us to {\it observe} the role of strong coupling effects and {\it perform} numerical simulations that non perurbatively handle strong coupling effects  {\it at the  cost of a weak coupling treatment}.
After introducing the  RCPT method, rather than focusing on benchmarking it against numerically-exact approaches, we exemplify the physics revealed by the formalism  and predict signatures of strong coupling in several classes of open quantum systems. 
We apply the RCPT approach and tackle five central classes of problems encountered in quantum transport and quantum thermodynamics. Namely, we study strong system-bath coupling effects in (a) thermalization, (b) thermal energy transport, 
(c) refrigeration,  (d) phonon-coupled electronic transport in the context of 
thermoelectric power generators, and (e) many-body, dissipative spin lattice physics. 
Figure \ref{fig:diagram-models} illustrates these five models. 
The paper is organized as follows: 
We present the reaction coordinate polaron transformation theoretical framework in Section \ref{sec:general}. 
We exemplify the method on five key open-system problems and demonstrate that their behavior is greatly altered by strong system-bath  coupling effects:
quantum thermalization  (Section \ref{sec:qubit}),  steady state heat transport (Section \ref{sec:qubitheat}),
quantum refrigeration  (Section \ref{sec:QAR}), 
phonon-assisted charge transport in thermoelectric engines
(Section \ref{sec:thermoE}), 
and steady state dissipative spin-chain models (Section \ref{sec:chain}).
In  each Section, theoretical predictions from the RCPT method are illustrated by numerical examples.
 We discuss and summarize our findings in Section \ref{sec:conclusion}.

\section{The RCPT theoretical framework}
\label{sec:general}

In this section, we describe the protocol for transforming an open system Hamiltonian 
using the reaction coordinate-polaron transformation (RCPT), thus, 
arriving at what we refer to as an \textit{effective Hamiltonian},
allowing for the interpretation of strong coupling effects in open quantum systems at low  cost. For simplicity, we present here the approach assuming a single heat bath; generalizations are discussed in Sec. \ref{sec:methodE}.
We work in units of $\hbar \equiv1$, $k_B\equiv 1$ and $e\equiv 1$.

We consider a generic open quantum system model, with an impurity system coupled to a bosonic reservoir,
\bea
\hat{H} = \hat{H}_s + \sum_k \nu_k \left(\hat{c}_k^{\dagger} + \frac{t_k}{\nu_k}\hat{S} \right) \left(\hat{c}_k + \frac{t_k}{\nu_k}\hat{S} \right).
\label{eq:H}
\eea
In this expression, $\hat{H}_s$ is the Hamiltonian of the system. $\hat{S}$ is a system operator, and it couples to the displacement of  reservoir modes of frequency $\nu_k$ with coupling strength $t_k$ assumed to be a real number;
$k$ is a wavevector index. Furthermore, $\hat{c}^{\dagger}_k$ ($\hat{c}_k$) are the bosonic creation (annihilation) operators for the bath modes. The interaction between the system and the reservoir is fully captured by the spectral density function, $J(\omega) = \sum_k t_k^2 \delta(\omega-\nu_k)$. 

\begin{figure}
\centering
\includegraphics[width=1\columnwidth]{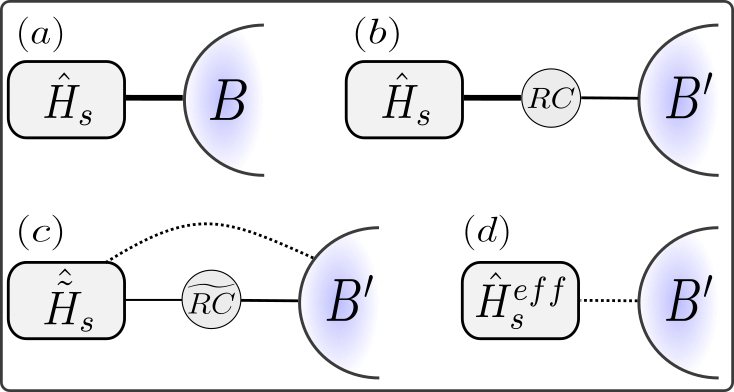}
\caption{Diagrammatic representation of the RCPT Hamiltonian transformations and truncation:
(a) The 
original model: An open system ($\hat H_s$) coupled to a bosonic bath ($B$). 
 (b) The system is extended to include a reaction coordinate, which is extracted from the reservoir, leaving a residual bath $B'$ missing one mode. 
 (c) The model after the application of the polaron transformation: The rotated system's Hamiltonian connects to both the RC and the residual bath. 
 (d) The model after the truncation of the RC, resulting in a so-called effective Hamiltonian with the system coupled only to the residual bath. This model is reminiscent of the original system in step (a).}
\label{fig:diagram}
\end{figure}

The framework consists of three steps, illustrated in Fig. \ref{fig:diagram}:
(i) An exact reaction coordinate mapping is performed on the bosonic reservoir, identifying a central degree of freedom to be extracted from the reservoir and incorporated as part of the system. This creates an \textit{extended} open system, which comprises the  original system along with its coupled reaction coordinate mode. This extended system is coupled to the residual bath with a modified spectral density function, typically with a weakened coupling strength compared to the original model, Eq. (\ref{eq:H}).
(ii) A polaron transformation  is applied
on the reaction coordinate. The transformation ``imprints" features of the RC into the original system, and partially decouples the RC from the system. This step further generates new {\it direct} interaction terms between the original system and the residual bath, which  provides insight into strong coupling features of the model. 
The transformations (i) and (ii)  are exact and unitary. 
(iii) The Hamiltonian is truncated
assuming that only the ground state of the reaction coordinate is populated. 
This approximate step relies on the reaction coordinate frequency (which derives from the original bath's spectral structure)
being the largest energy scale in the problem, notably, exceeding the thermal energy, which is our working assumption here. More details on the consequences of this approximation are included in Sec. \ref{sec:Approximation}.

We refer to the application of these three steps in succession as the RCPT method. Once the RCPT procedure is performed, an effective Hamiltonian emerges, which mathematically resembles the original model, Eq. (\ref{eq:H}). 
However, parameters in the new, effective model contain an explicit dependency on the original system-bath coupling parameters. This in turn allows for the interpretation of strong coupling features. In what follows, we present this protocol to generate effective Hamiltonian models as a means of capturing strong coupling effects in open quantum systems. 

\subsection{Reaction coordinate mapping}
\label{sec:RC1}

The first step in deriving effective Hamiltonian models using the RCPT method is to perform an exact reaction coordinate mapping \cite{Nazir18} on the Hamiltonian Eq. (\ref{eq:H}). This transformation results in the extraction of a collective reservoir mode of frequency $\Omega$, which couples to the system at strength $\lambda$, and is included as part of the system. 
We refer to the resulting Hamiltonian as the \textit{reaction coordinate Hamiltonian},
\bea \label{eq:HRC}
\hat{H}_{RC} &= &\hat{H}_s +  \Omega \left(\hat{a}^{\dagger} + \frac{\lambda}{\Omega}\hat{S} \right) \left(\hat{a} + \frac{\lambda}{\Omega}\hat{S} \right) \nonumber 
\\
&+& \sum_k \omega_k \left(\hat{b}_k^{\dagger} + \frac{f_k}{\omega_k} (\hat{a}^{\dagger} + \hat{a}) \right) \left(\hat{b}_k + \frac{f_k}{\omega_k} (\hat{a}^{\dagger} + \hat{a}) \right),
\eea
where the reaction coordinate is defined such that 
\bea
\lambda (\hat{a}^{\dagger}+\hat{a}) = \sum_k t_k (\hat{c}^{\dagger}_k+\hat{c}_k).
\eea
In the above expression, 
the bosonic creation (annihilation) operator of the RC is $\hat{a}^{\dagger}$ ($\hat{a}$) and it is coupled with strength $f_k$
to the residual bath, identified by the creation  (annihilation)  bosonic operators $\hat{b}^{\dagger}_k$ ($\hat{b}_k$)
 of frequency $\omega_k$.
In the RC representation, 
the coupling parameter $\lambda$ between the system and the reaction coordinate, and the frequency of the reaction coordinate $\Omega$ are obtained from the original spectral density function via the expressions \cite{Nazir18}
\bea
\lambda^2 = \frac{1}{\Omega} \int_0^\infty \omega J(\omega) d\omega,
\eea
and
%
\bea
\Omega^2 = \frac{\int_0^\infty \omega^3 J(\omega) d\omega}{\int_0^\infty \omega J(\omega) d\omega}.
\eea
Note that $\lambda$ characterizes the extent of interaction between the original system and the bath.
As such, it is a central parameter to tune in the exploration of quantum dissipative behavior at strong coupling.

In the RC picture, 
a different spectral density function now characterizes the interaction between the RC and the residual bosonic environment, 
$J_{RC}(\omega) = \sum_k f_k^2 \delta(\omega - \omega_k)$. 
From Heisenberg's equation of motion, it can be shown that this newly-defined spectral density function is related to the original spectral density function by\cite{Nazir18}
\bea
J_{RC}(\omega) = \frac{2\pi \lambda^2 J(\omega)}{\left[ P \int \frac{J(\omega') d\omega'}{\omega' - \omega} \right]^2 + \pi^2J(\omega)^2},
\label{eq:JRC}
\eea
where in this expression, the integration is understood as a principal-value integral. 

In the reaction coordinate representation, the system-reservoir boundary is shifted to include an additional mode from the reservoir, resulting in an extended system, given by the first two terms of Eq. (\ref{eq:HRC}). The last term of Eq. (\ref{eq:HRC}) represents the residual environment and its interaction to the reaction coordinate.

The power of the RC transformation stems from the fact that  when increasing the coupling strength, $J(\omega)\to
\alpha J(\omega)$ with $\alpha > 0$, only the system-reaction coordinate coupling strength gets modified to $\lambda \to \sqrt{\alpha} \lambda$, while $J_{RC}(\omega)$ does not change with $\alpha$. 
This allows one to perform perturbative quantum master equation calculations on the enlarged system, providing a valid strong coupling treatment relying on weak coupling tools. 
%




\subsection{Polaron transformation}             \label{sec:RC2}

In the second step of the RCPT method we perform a polaron transformation on the reaction coordinate Hamiltonian, Eq. (\ref{eq:HRC}).
%
This unitary transformation is given by the shift operator
\bea
\hat{U}_P = e^{\frac{\lambda}{\Omega} (\hat{a}^{\dagger} - \hat{a})\hat{S}},
\eea
which partially decouples the system from the RC, as we explain below. 
As a consequence of this transformation, we generate new {\it direct} coupling terms between the residual bath and the system, as well as the RC and the system. 
We note that our polaron operator lives in the Hilbert space of the system and the RC. 
%
To perform the transformation, we rely on the fact that
$\hat{U}_P \hat{a} \hat{U}_P^{\dagger} = \hat{a} - \frac{\lambda}{\Omega}\hat{S}$,
and use the shorthand notation for the polaron-transformed reaction coordinate system Hamiltonian 
$\hat{\tilde{H}}_s\equiv \hat{U}_P \hat{H}_s \hat{U}_P^{\dagger}$. The total polaron-transformed reaction coordinate Hamiltonian is $\hat{H}_{RC-P} \equiv \hat{U}_P \hat{H}_{RC} \hat{U}_P^{\dagger}$,
 given by
\bea \label{eq:HRCP}
\hat{H}_{RC-P} &=& \hat{\tilde{H}}_{s} + \Omega \hat{a}^{\dagger}\hat{a} \nonumber
\\
&&+\sum_k \omega_k \bigg\{\left[ \hat{b}_k^{\dagger} + \frac{f_k}{\omega_k} \left(\hat{a}^{\dagger} + \hat{a} -\frac{2\lambda}{\Omega}\hat{S}\right) \right]\nonumber\\ 
&&\;\;\;\;\;\;\;\;\;\;\;\;\times\left[ \hat{b}_k + \frac{f_k}{\omega_k} \left(\hat{a}^{\dagger} + \hat{a} -\frac{2\lambda}{\Omega}\hat{S}\right) \right]\bigg\}.
\nonumber\\
\eea
%
Since the system Hamiltonian is dressed by the polaron transformation operator, it is now a function of the system-bath coupling parameter $\lambda$. Another way to think about this, is that the transformation ``imprints" the RC into the system's Hamiltonian.
Furthermore, new interaction terms emerge in the polaron frame. Namely, the system now couples {\it directly} to both the RC ($\hat a$ operators)  and to the residual bath ($\hat b_k$ operators).
However, the functional form of the spectral density function of the residual bath is  unaltered by this transformation, and it is still captured by Eq. (\ref{eq:JRC}), albeit with an additional prefactor $(2\lambda/\Omega)^2$.

At first glance, applying the polaron operator after the RC transformation seems unconducive for performing calculations, as there are now terms coupling the system to both the RC and the residual environment. In fact, we appear to have made the Hamiltonian more complex to solve due to the addition of new interaction terms.
As we show next, the Hamiltonian Eq. (\ref{eq:HRCP}) is actually an intermediate step in deriving effective Hamiltonian models and after
an additional simplifying approximation, it becomes tractable.
%

\subsection{Reaction coordinate truncation}
\label{sec:RC3}

The RC and the polaron transformations are exact. As such, no approximations have been made up to this point. 
To simplify the Hamiltonian (\ref{eq:HRCP})
we now invoke an approximation, generating an \textit{effective Hamiltonian} $H^{eff}(\lambda)$. This effective Hamiltonian
is transparent for analytical work and it also serves as a good starting point for numerical implementations.

Assuming $\Omega$, the frequency of the RC, to be the largest energy scale in the problem (see a discussion in Sec. \ref{sec:Approximation}) ---
higher in particular than the temperature of the attached bath(s), $\Omega \gg T$---
one can safely truncate the harmonic manifold of the RC and consider only its ground level.
The truncation does not eliminate strong coupling effects in a regime where $\Omega$ is the largest energy scale.
%
We thus define the effective Hamiltonian as the subspace of Eq. (\ref{eq:HRCP}) in which the RC is fixed to its ground state $|0\rangle$,
\bea
\hat{H}^{eff}(\lambda) = \bra{0}\hat{H}_{RC-P}\ket{0}.
\label{eq:Heff1}
\eea
Explicitly,
\bea
\hat{H}^{eff}(\lambda) &= &\bra{0} \hat{\tilde{H}}_{s} \ket{0} 
 \nonumber\\
&+& \sum_k \omega_k \left(\hat{b}_k^{\dagger} - \frac{2 \lambda f_k}{\Omega \omega_k}\hat{S} \right) \left(\hat{b}_k - \frac{2  \lambda f_k}{\Omega \omega_k}\hat{S} \right).
\label{eq:Heff}
\eea
We highlight here the dependency of the effective system Hamiltonian on the coupling parameter $\lambda$; 
however, the effective model  depends as well on additional system parameters, such as the RC frequency, $\Omega$.

Crucially, the effective Hamiltonian (\ref{eq:Heff}) has a mathematical structure similar to the original Hamiltonian, Eq. (\ref{eq:H}). Two important distinctions are, however, apparent: (i) The system Hamiltonian is dressed by the polaron transformation operator,
%
\bea
\hat{H}_s^{eff}(\lambda) &=& \bra{0} \hat{\tilde{H}}_{s}
\ket{0}.
\eea
(ii) The reservoir considered here is the residual bath that was obtained in the RC mapping, Eq. (\ref{eq:HRC}). The spectral density function of the bath is further dressed by the RC parameters,
\bea
J^{eff}(\omega) &=& \frac{4\lambda^2}{\Omega^2} J_{RC}(\omega).
\label{eq:Jeff}
\eea
This completes the introduction of the RCPT method. Considering a general Hamiltonian strongly coupled to a bath, Eq. (\ref{eq:H}),
we mapped the model into Eq. (\ref{eq:Heff}), with a $\lambda$-dressed system Hamiltonian and a {\it weak} system-bath coupling. 
This is because  $J_{RC}(\omega)$ describes the coupling of the extended system to a residual bath, which is weakened relative to the original case. As well, we further assume that $\lambda/\Omega\ll 1$ for $\Omega$ to retain its status as the largest energy scale in the problem.

The RCPT method allows nonpertubative studies of strong coupling models.
The elegant and powerful aspects of the RCPT method stem from the effective Hamiltonian (\ref{eq:Heff}) seemingly having the same complexity as the original model. This in turn allows the effective Hamiltonian to be treated with weak coupling methods since the original system-bath coupling energy $\lambda$ has been quenched and absorbed into redefining the system itself.
The main novelty  of the method lies in it allowing analytical studies of the strong coupling regime. 
Since at weak coupling one can often approach the problem analytically and acquire closed-form expressions
for e.g., nonequilibrium steady state properties, one can now build on these solutions---only with renormalized parameters and a dressed system Hamiltonian. 
Moreover, even without performing a detailed analysis, the form of the Hamiltonian (\ref{eq:Heff}), compared to Eq. (\ref{eq:H}), immediately exposes contributions of strong coupling to the open quantum system, e.g., in opening new transport pathways and shifting parameters.

We now go one step further and manipulate the effective system Hamiltonian to cast it in a more transparent form, which will be useful in applications,
\bea
\hat{H}_{s}^{eff}(\lambda) = \bra{0}e^{\frac{\lambda}{\Omega} (\hat{a}^{\dagger} - \hat{a})\hat{S}} \hat{H}_{s} e^{-\frac{\lambda}{\Omega} (\hat{a}^{\dagger} - \hat{a})\hat{S}} \ket{0}.
\label{eq:HSeff0}
\eea
The polaron operators are mathematically analogous to displacement operators encountered in quantum optics. They have the useful property of generating coherent states when acting on the vacuum $D(\alpha)\ket{0} = \ket{\alpha}$. Furthermore, coherent states may be written as a superposition of harmonic oscillator eigenstates. We employ these two properties defining $\alpha \equiv \frac{\lambda}{\Omega}\hat{S}$ as an operator. This results in a useful form for the effective system Hamiltonian,
\bea
\hat{H}_s^{eff}(\lambda) = e^{-\frac{\lambda^2}{2\Omega^2} \hat{S}^2}\left( \sum_{n=0}^\infty \frac{\lambda^{2n}}{\Omega^{2n} n!} \hat{S}^n\hat{H}_s \hat{S}^n \right) e^{-\frac{\lambda^2}{2\Omega^2} \hat{S}^2}.
\label{eq:HSeff}
\eea
This expression allows for the computation of the effective system Hamiltonian, which is now coupling-strength dependent. 
Further details are given in Appendix A. 

The effective Hamiltonian, Eq. (\ref{eq:Heff}), concludes the theoretical account of the RCPT framework. 
In Sec. \ref{sec:spectral}, we summarize the evolution of the spectral density function during the RCPT steps.
We discuss the assumptions of the RCPT method and thus its regime of validity in Sec. \ref{sec:Approximation}.
Theoretical extensions to the basic framework are presented in Sec. \ref{sec:methodE}.
The numerical QME implementation of the method to study transport behavior is described in Sec. 
\ref{sec:Redfield}.


\subsection{Evolution of the spectral density function in the RCPT method}
\label{sec:spectral}

The RCPT method is not limited to a specific type of spectral density function and the procedure outlined above is general.
However, to make the method useful, one should work in a parameter range such that, though the original model may carry strong couplings to the bath,  $J_{RC}(\omega)$ corresponds to a weak-coupling situation.
In this work, we exemplify the RCPT method using the Brownian spectral density function,
\bea
J(\omega) = \frac{4\gamma \Omega^2 \lambda^2 \omega}{(\omega^2 - \Omega^2)^2 + (2\pi\gamma \Omega \omega)^2}.
\label{eq:Brownian}
\eea
In this model, the system's coupling to the bath is peaked at $\Omega$ with a width parameter $\gamma$.  
$\lambda$ tunes the system-reservoir coupling strength.
It can be shown \cite{NazirPRA14,Nazir16,Nick2021} that performing a reaction coordinate mapping  translates, via  Eq. (\ref{eq:JRC}),
this spectral density function to an Ohmic form
\bea
    J_{RC}(\omega) = \gamma \omega e^{-|\omega|/\Lambda}.
\label{eq:Ohmic}
\eea
This expression becomes exact when $\Lambda$, the cutoff energy, tends to infinity. 
In the RC representation, the dimensionless width parameter $\gamma$ controls the coupling strength between the RC and the residual environment. Furthermore, in this model the location of the central peak, $\Omega$, maps to the frequency of the reaction coordinate.
Thus, a narrow Brownian function translates to an extended system weakly coupled to the residual bath. 

After the polaron transformation we build the effective model and the spectral density function is further dressed 
according to Eq. (\ref{eq:Jeff}), ending with
\bea
    J^{eff}(\omega) = \frac{4\lambda^2}{\Omega^2}\times \gamma \omega e^{-|\omega|/\Lambda}.
    \label{eq:Ohmiceff}
\eea
%
In what follows, we describe bosonic baths using the Brownian spectral density function. In contrast, 
we treat fermionic reservoirs under the weak system-reservoir coupling assumption and take them in the wideband limit.

%

\subsection{Regimes of applicability of the RCPT formalism}
\label{sec:Approximation}

In this section, we discuss the only approximation of the RCPT mapping: representing the manifold of the RC solely by its ground state. 
We first identify the parameter regime where this approximation holds. We then discuss 
the type of problems that would benefit from being addressed by the RCPT technique. 
We emphasize that once the RCPT mapping is complete, and we reach the effective Hamiltonian Eq. (\ref{eq:Heff}), different analytical and numerical tools can be implemented on the effective Hamiltonian. These tools may come with their own independent sets of approximations; in this section however, we focus exclusively on the approximation of the RCPT mapping.

Considering an open quantum system model,
we list the relevant energy scales in the problem:
$\Delta$ would serve as a characteristic energy scale for the system, e.g. spin splitting in a spin-bath model;
$\Omega$ is a characteristic frequency for the bath; $\lambda$ characterizes the system-bath coupling; $T$ is the thermal energy. 

Let us now discuss the energy spectrum of the RC Hamiltonian: 
For a large value of $\Omega$ relative to the eigenspectrum of the bare system and the coupling energy, $\Omega\gg \Delta$, $\lambda$,
the energy spectrum of Eq. (\ref{eq:HRC})
shows manifolds of levels separated by gaps of ${\cal O}(\Omega)$. In each manifold, levels roughly correspond to the original system thus they are spaced by ${\cal O}(\Delta)$.
%
The truncation of the RC performed in Sec. \ref{sec:RC3} is justified as long as 
$\Omega\gg T$:
In this regime, thermal energy from the bath is insufficient to significantly populate higher excited states of the reaction coordinate. 

Altogether, the RCPT formalism presented in this work is expected to be valid when $\Omega\gg \Delta,\lambda$, and 
$\Omega\gg T$. Note that there is no limitation on whether the thermal energy is higher or lower than characteristic energies in the system, $\Delta$.
In impurity models, the temperature is defined relative to the eigenvalues of the system Hamiltonian. Therefore, the RCPT procedure is valid for both high and low temperatures.




What type of problems would benefit from representing them with the effective Hamiltonian? %
The truncation of the RC to its ground state 
accurately captures the impact of strong coupling in transport phenomena, as we show in the next Sections.
However, the truncation drastically curtails the ability to follow dynamical effects, as we now explain.
The reaction coordinate method captures non-Markovian dynamics, see e.g., Refs. \cite{NazirPRA14,Nickdecoh}. This is possible because of the build-up of correlations during time evolution between the system and the RC (which in truth is part of the bath). 
An undesired consequence of the RC truncation is losing this ability to exchange
information between the RC and the system, hence, missing dynamical features that emerge due to non-Markovianity. In other words, in the RCPT formalism, the RC does not evolve in time; it is maintained in its ground state.
%
As a result, transient features in the dynamics, e.g., some oscillations, would be missed.
However, the RCPT method recovers the correct decay constants at strong couplings, thus
the steady state limit is well described. 

The RCPT method can be systematically made more accurate by keeping higher excited states of the RC. This approach, described in Appendix B would recover missing dynamical features.
However, this is achieved at the cost of exponentially increasing the dimensionality of the Hilbert space of the system, thus losing insights gained from the elegant effective Hamiltonian picture.

\subsection{Extensions}
\label{sec:methodE}

\subsubsection{Iterative mapping}
In implementing the RCPT method on Eq. (\ref{eq:H}) we arrived at a form that closely resembles the initial Hamiltonian, except with a $\lambda$-dressed effective system Hamiltonian and a different spectral density function. 
In principle, it should be possible to iterate this process: repeatedly extract an RC mode from the bath,
perform a polaron transformation on this mode, and truncate the RC to only occupy its ground state.
This process would lead to an effective description including strong coupling effects
even with highly-structured spectral density functions, for example, a bimodal function.
The effective Hamiltonian after $n$ such rounds (where $n$ is still significantly smaller than the number of modes in the bath) 
would have the following structure
\bea
\hat{H}^{eff}_{n} = && \bra{\bold{0}_n} \hat{\tilde{H}}_s^{(n)} \ket{\bold{0}_n} \nonumber
\\ 
&&+  \sum_k  \omega_k \Bigg \{ \left[\hat{b}_k^{\dagger} + \prod_{i=1}^{n} \left ( (-1)^i \frac{2\lambda_i}{\Omega_i} \right )\frac{f_k}{\omega_k} \hat{S} \right ] \nonumber
\\ 
&&\;\;\;\;\;\;\;\;\;\;\;\;\times\left [ \hat{b}_k + \prod_{i=1}^{n} \left(  (-1)^i \frac{2\lambda_i}{\Omega_i} \right ) \frac{f_k}{\omega_k} \hat{S} \right ] \Bigg \} .
\eea
In this expression, we used a shorthand notation to express the effective system Hamiltonian,
\bea
\bra{\bold{0}_n} \hat{\tilde{H}}_s^{(n)} \ket{\bold{0}_n} \equiv \bra{0_1,...,0_n} \prod_{i=1}^n (\hat{U}_{P,i}) \hat{H}_s \prod_{i=1}^n (\hat{U}^{\dagger}_{P,i}) \ket{0_1,...,0_n},
\nonumber\\
\eea
with $|0_n\rangle$ denoting the ground state of the $n$th RC.
The evaluation of this term is in fact simple since the sequence of polaron transformations commute. 
Additionally, the spectral density functions can be iteratively computed using Eq. (\ref{eq:JRC}). 
It is therefore straightforward to iterate this process as desired. 
We emphasize that even without iterating, that is following the procedure \ref{sec:RC1}-\ref{sec:RC3}, the Hamiltonian includes strong coupling effects in a nonperturbative manner.

\subsubsection{Multibath problems}

Another extension of interest concerns the application of this tool to study open quantum systems coupled to multiple environments. 
It is straightforward to extract simultaneously more than one RC, e.g., one from each bath.  
However, the polaron transform can become quite complicated when the extended model includes more than one RC. 
Namely, it is not guaranteed that the individual 
polaron transformations on each RC will commute with one another, and this aspect depends on the details of the model.
As such, we are left with the arbitrary freedom of deciding which polaron transform to apply first, potentially changing the outcome of the calculation (a fact that is not surprising, as this is not an exact tool). 
Given this non-uniqueness of the procedure, one would need to test different sequences of the polaron transformation
 and  select the the most feasible and tractable approach.
 
 In summary, the RCPT approach is straightforward to apply in situations in which the system's operators that couple to the different baths commute, or, if one applies only a single polaron transformation before truncating the reaction coordinate.

\subsection{Numerical implementation: Redfield QME}
\label{sec:Redfield}

Before tackling examples of impurity models with the RCPT method,  we briefly review the numerical approach used in this work to simulate
the steady state limit of the reduced system's dynamics, as well as different currents.
We implement the Redfield quantum master equation (QME), which relies on weak coupling and Markov approximations, and simulate the nonequilibrium behavior of three Hamiltonians:

(i) The original model, Eq. (\ref{eq:H}). 


(ii) The reaction coordinate Hamiltonian, Eq. (\ref{eq:HRC}). 

(iii) The effective Hamiltonian, Eq. (\ref{eq:Heff}), which constitutes the last step of the RCPT method.

We emphasize that in all cases we use the Born-Markov Redfield (BMR)
quantum master equation, which in case (ii) is performed on the extended system, 
and in (iii) on the effective model. 
In both of these cases this second-order 
method is able to capture strong coupling effects. In contrast, in (i) the BMR-QME provides meaningful results only if the coupling strength in the original picture is weak.
We first discuss the general Redfield equation, and then comment on modifications required to study each variation, (i), (ii) and (iii). 
Working in the Schr\"odinger representation and in the energy basis of the system Hamiltonian, the Redfield equation for the reduced density matrix $\rho(t)$ of the system is given by
\bea
\dot{\rho}_{mn}(t) &=& -i\omega_{mn} \rho_{mn}(t)
\nonumber\\ &-&
\sum_{j,p}[
R_{mj,jp}(\omega_{pj}) \rho_{pn}(t) + R_{np,pj}^{*}(\omega_{jp}) \rho_{mj}(t)
\nonumber\\ &-&
R_{pn,mj}(\omega_{jm}) \rho_{jp}(t) - R_{jm,np}^{*}(\omega_{pn}) \rho_{jp}(t)].
\label{eq: Redfield}
\eea
The indices $m$  (as well as $n$, $j$ and $p$) denote eigenstates of the system with eigenvalues $E_{m}$, and Bohr frequencies
$\omega_{mn}\equiv E_m-E_n$. 
The elements of the $R$ superoperator are given by a half Fourier transform of bath autocorrelation functions,
\bea
R_{mn,jp}(\omega) &=& (S^{D})_{mn} (S^{D})_{jp} \int_0^{\infty} d\tau e^{i\omega\tau} \langle \hat{B}(\tau)\hat{B}(0)\rangle
\nonumber\\ &=&
(S^{D})_{mn} (S^{D})_{jp} [\Gamma(\omega) + i\Delta(\omega)],
\label{eq:dissipator}
\eea
with $\hat{S}^D$ standing for the system's operator that is coupled to the bath, 
written in the energy basis of the system Hamiltonian. 
Furthermore, $\Gamma(\omega)$ and $\Delta(\omega)$ are the real and imaginary parts of
 the bath autocorrelation function, respectively. These correlation functions  
are evaluated with respect to the thermal state of the bath. 
In this work, we neglect the imaginary part of the autocorrelation function as it contributes only a small shift to the spectrum.

For harmonic environments and a bilinear system-bath coupling, the real part of the $R$ tensor evaluates to
\bea
\Gamma(\omega) =
\begin{cases}
\pi J(|\omega|) n(|\omega|) & \omega < 0 ,  \\
\pi J(\omega)[n(\omega) + 1] & \omega > 0, \\
\pi \lim_{\omega \to 0} J(\omega) n(\omega)  & \omega = 0.
\end{cases}
\eea
Here, $n(\omega)$ is the Bose-Einstein distribution 
function of the bath, characterized by an inverse temperature $\beta=1/T$. 
In a compact form, the evolution of the system density matrix is given by
\bea
    \dot{\rho}(t) = -i[\hat{H_s},\rho(t)] + \sum_\alpha D_{\alpha}(\rho(t)),
\eea
where we have already generalized the equation to include 
multiple thermal reservoirs with the dissipators $D_{\alpha}(\rho(t))$, 
organized based on Eq. (\ref{eq: Redfield}). 

We solve the equation of motion in the steady state,  and obtain the density matrix of the system, $\rho^{SS}$.
This can be achieved in different ways; here  we write down the equation of motion in a compact form as 
$\dot{\rho}(t) = {\mathcal L}\rho(t)$ and we further construct a modified Liouvillian $\mathcal L'$ by replacing the last row with a population (probability) conservation condition. We also define the column vector $v$ with all its elements set to zero besides the last one, which corresponds to the population conservation condition with the diagonal elements of the system's density matrix summing up to unity.
The steady state limit of the system's density matrix is then obtained by algebraic operations,
\bea
 \mathcal L' \rho^{SS} = v.
 \label{eq:SSpop}
\eea
This formalism allows the calculation of currents;
the heat current at the $\alpha$th contact, for example, is
calculated from the heat exchanged between the system and the $\alpha$th reservoir,
\bea
    j_{q}^{\alpha}(t) = \Tr_s\left[ D_{\alpha}(\rho(t))\hat{H}_s \right],
\eea
where steady state currents are obtained once obtaining the state of the system in the long time limit, $\rho^{SS}$.
The heat current is defined positive when flowing from the $\alpha$th bath towards the system. 
Similarly, the charge current at the $\alpha$th contact is
\bea
 j_{e}^{\alpha}(t) = \Tr_s\left[ D_{\alpha}(\rho(t))\hat{N_s} \right],
 \label{eq:curre}
\eea
where $\hat N_s$ is the number operator for the system.

We now elaborate on the three implementations of the Redfield QME that we use in this work:

(i) BMR-QME simulations refer to using the
Born-Markov Redfield quantum master equation
directly on  the original Hamiltonian, Eq. (\ref{eq:H}), with the associated spectral density function of the heat 
bath, here Eq. (\ref{eq:Brownian}).
The BMR-QME provides inaccurate results 
in the strong system-bath coupling regime, as the method is stretched beyond its regime of validity. 
In this sense, we regard the BMR-QME as an asymptotic solution of the weak coupling limit.

(ii) RC-QME simulations refer to using the
Redfield QME to study the 
extended system after adding the RC, Eq. (\ref{eq:HRC}), 
with the relevant spectral density function, here Eq. (\ref{eq:Ohmic}) (or more generally Eq. (\ref{eq:JRC})). 
In practice, we  truncate the RC harmonic oscillator manifold, 
reducing it to its first $M$ levels. 
The coefficients $(S^D)_{mn}$ in the dissipator, Eq. (\ref{eq:dissipator}),
are dictated by the form of the RC coupling to the bath, see the third term in Eq. (\ref{eq:HRC}). 
 While the original system may be strongly coupled to the bath,
in the RC representation the assumption of a weak coupling between the {\it residual} reservoir and the {\it extended} system can be 
justified as explained in Sec. \ref{sec:spectral}.


(iii) EFF-QME simulations correspond to
using the Redfield QME, except we now apply it on the effective Hamiltonian,
Eq. (\ref{eq:Heff}) with the spectral density function,  Eqs. (\ref{eq:Jeff}) and (\ref{eq:Ohmiceff}). 
Recall that the effective Hamiltonian is constructed with the RCPT framework.
This approach allows for strong coupling effects to be captured through renormalized 
parameters, while still retaining a simple and tractable quantum master equation framework;
the dimension of the Hilbert space of the effective system is equal to the original model.
This approach is in principle less accurate than the RC-QME since the RC manifold is truncated. 
However, the EFF-QME method should reliably capture predictions of the RC-QME in the 
limit where $\Omega$ is the largest energy scale, see Sec. \ref{sec:Approximation} for a discussion on this point.

The hierarchy of methods goes as follows: 
The RC-QME provides the most accurate results at strong coupling, maintaining dynamical effects in the RC. 
However, it is a numerical method and it does not offer deep insights into the physical mechanisms of strong coupling.
The EFF-QME method preserves dominant strong coupling effects, but is less accurate. On the other hand, it provides a profound understanding of underlying strong coupling effects.
The BMR-QME is valid only at weak system-bath coupling.

In what follows, we study and simulate several prominent impurity models with the BMR-QME, RC-QME, and EFF-QME methods
to illustrate the predictive power of the RCPT approach. Our main argument however is
that the effective Hamiltonian itself --- namely Eq. (\ref{eq:Heff}), the outcome of the RCPT approach --- already 
allows assessment of contributions of strong coupling to transport characteristics, even {\it without} performing simulations.


\section{Thermalization: Weak, intermediate and the ultrastrong coupling limit}
\label{sec:qubit}

What is the equilibrium state of a system that is coupled to a heat bath at temperature $T$? For macroscopic objects,
statistical physics asserts that in the long time (steady state) limit, the system should reach the conventional Gibbs state, $\rho^{SS}=\frac{1}{Z}e^{-\beta\hat H_s}$,
with $\hat H_s$ the Hamiltonian of the system, $\beta$ the inverse temperature of the bath and $Z={\rm Tr}_s\left[e^{-\beta \hat H_s}\right]$ the relevant partition function, with the trace performed over the system's degrees of freedom. 
However, the Gibbs state assumption is valid only if the interaction of the system with the heat bath is vanishingly weak;
 it  breaks down, e.g., for nanoscale systems once the interaction energy becomes comparable to energy parameters of the system. 
The derivation of the equilibrium state as a function of the system-bath interaction energy has been a topic of recent focus, particularly when steady state coherences are generated \cite{Giacomo,Archak,Valkunas20,Cresser,Zhang_2022,Trusheckin_2022}.
Note that in this Section the system is  coupled to a single heat bath, and we thus refer to the equilibrium state as the steady state. 
In the next Sections, when dealing with multiple heat baths, the steady state is a nonequilibrium state. 

What is then the long-time state of a quantum system coupled to a heat bath?  The general statement is that the system should reach the mean-force Gibbs state (MFGS), defined as
\bea
\rho^{SS}_{MFGS}= \frac{1}{Z_{MFGS}}{\rm Tr}_B\left [e^{-\beta \hat H}\right],
\eea
which is  the state obtained once taking a  partial trace over the bath's degrees of freedom.
Here, $\hat H$ is the total Hamiltonian and the partition function is defined with the the full trace, $Z_{MFGS}={\rm Tr}[e^{-\beta \hat H}]$.
Generally, this MFGS {\it differs} from the conventional Gibbs state. An analytic expression for the MFGS was obtained
for the Caldeira-Leggett model in the ultra-strong coupling limit \cite{Cresser,Trusheckin_2022}. It was shown that 
in this case, the equilibrium state of the system was diagonal --- albeit in the basis of the system's operator that couples to the bath.

\begin{figure*}[htbp]
    \centering
    \includegraphics[width=2\columnwidth]{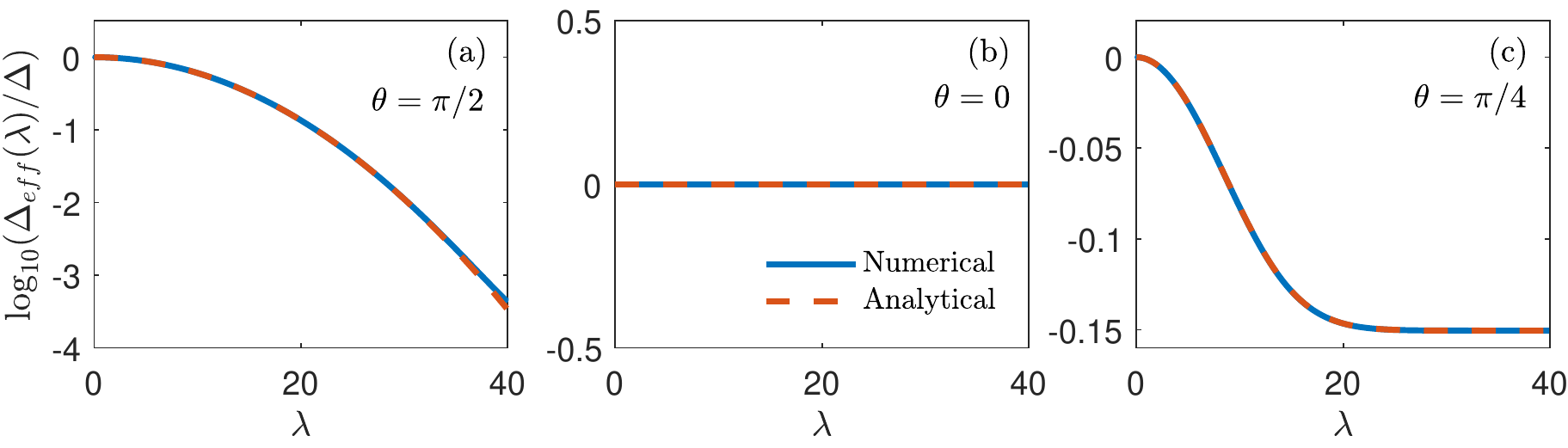}
\caption{Spectrum of the generalized spin boson model.  
We display $\Delta_{eff}(\lambda)$, the energy gap between the first excited and the ground state eigenenergies of the spin system as a function of the system-bath coupling for three different angles: (a) $\theta=\pi/2$, (b) $\theta=0$, and (c) $\theta=\pi/4$. Analytical expressions obtained from the effective spin splittings Eqs. (\ref{eq:deltapi0})-(\ref{eq:deltapi4}) (dashed) perfectly agree with results from the numerical diagonalization of Eq. (\ref{eq:HRC}), applied to the generalized spin-boson model (full). Parameters used here are $\Delta=1$ and $\Omega = 20$.}
    \label{fig:Spectrum}
\end{figure*}

Here we show that the effective model, the outcome of the RCPT process, provides an excellent {\it analytic} approximation for the MFGS from weak coupling, through the intermediate regime, to the ultrastrong coupling limit. 
We perform this analysis on the generalized spin-boson model with a spin coupled to a harmonic reservoir,
%
%
\bea
\hat{H} = \Delta \hat{\sigma}_z + \sum_k \nu_k \left( \hat{c}_k^{\dagger} + \frac{t_k}{\nu_k}\hat{\sigma}_\theta \right) \left( \hat{c}_k + \frac{t_k}{\nu_k}\hat{\sigma}_\theta \right),
\label{eq:HSB}
\eea
where $\Delta$ is the spin splitting.
The spin is coupled to a bosonic bath with 
parameters as defined in Eq. (\ref{eq:H}). The system interaction operator to the bath is 
$\hat{\sigma}_{\theta} = \cos(\theta)\hat{\sigma}_z +  \sin(\theta)\hat{\sigma}_x$, where $0 \leq \theta \leq \pi/2$ is the angle of a vector pointing in the $x-z$ plane of the Bloch sphere, determining the nature of the system-bath interaction.
With this definition, $\theta = \pi/2$ corresponds to the standard spin-boson model while $\theta=0$ is the pure dephasing model where only decoherence dynamics is observed. 
In the language of Eq. (\ref{eq:H}), 
the generalized spin-boson model 
(\ref{eq:HSB}) corresponds to the choice
$\hat H_s=\Delta \hat \sigma_z$ and 
$\hat S=\hat \sigma_{\theta}$.

Based on Eqs. (\ref{eq:Heff})-(\ref{eq:HSeff}), we write down directly the effective Hamiltonian obtained from Eq. (\ref{eq:HSB}) through the RCPT procedure. 
Making use of the properties of the Pauli operators, namely, $\hat{S}^2 = 1$ and $\hat{S^3} = \hat{S}$, we break down the sum in Eq. (\ref{eq:HSeff}) into even and odd contributions,
\bea
\nonumber
\hat{H}_s^{eff}(\lambda) = e^{-\frac{\lambda^2}{\Omega^2}} \left( \sum_{n;even} \frac{\lambda^{2n}}{\Omega^{2n} n!} \hat{H}_s + \sum_{n;odd} \frac{\lambda^{2n}}{\Omega^{2n} n!} \hat{S}\hat{H}_s\hat{S} \right).
\nonumber\\
\eea
We sum the series noting that $\hat{S}\hat{H}_s\hat{S} = \Delta\sin(2\theta) \hat{\sigma}_x + \Delta\cos(2\theta)\hat{\sigma}_z$. The effective system Hamiltonian thus becomes
\bea
\hat{H}_s^{eff}(\lambda) &=&  e^{-\frac{\lambda^2}{\Omega^2}} \cosh(\frac{\lambda^2}{\Omega^2}) \Delta \hat{\sigma}_z 
\nonumber\\
&+& e^{-\frac{\lambda^2}{\Omega^2}} \left[\sinh(\frac{\lambda^2}{\Omega^2}) (\Delta \sin(2\theta) \hat{\sigma}_x + \Delta\cos(2\theta)\hat{\sigma}_z)\right].\nonumber\\ 
\eea
Next, we rearrange this expression into the form,
\bea
\nonumber
\hat{H}_s^{eff}(\lambda) &=& \frac{\Delta}{2}\left[(1+e^{-\frac{2\lambda^2}{\Omega^2}}) + (1-e^{-\frac{2\lambda^2}{\Omega^2}})\cos(2\theta)\right]\hat{\sigma}_z \nonumber\\ 
&+& \frac{\Delta}{2}\left(1-e^{-\frac{2\lambda^2}{\Omega^2}}\right)\sin(2\theta)\hat{\sigma}_x.
\label{eq:HSBeff}
\eea
This is the effective Hamiltonian for the system, and it uncovers two important aspects of strong coupling.
(i) Parameter renormalization: As can be seen from the first row in Eq. (\ref{eq:HSBeff}), the qubit splitting is suppressed when $\theta \neq 0$ once $\lambda>0$ since  $\frac{1}{2}\left[(1+e^{-\frac{2\lambda^2}{\Omega^2}}) + (1-e^{-\frac{2\lambda^2}{\Omega^2}})\cos(2\theta)\right]\leq 1$.
(ii) Generation of new processes: The second row in Eq. (\ref{eq:HSBeff}) reveals that a new system tunneling term appears for $0<\theta<\pi/2$, compared to the original Hamiltonian, Eq. (\ref{eq:HSB}). This term is induced by the system-bath coupling, $\lambda\neq0$.
To gain insight into the strong coupling features of this model, 
Eq. (\ref{eq:HSBeff}), we consider three angles as special cases:

(1) The pure-dephasing model is realized when $\theta = 0$. 
This reduces Eq. (\ref{eq:HSBeff}) to
\bea
\hat{H}_s^{eff}(\lambda,\theta=0) = \Delta \hat{\sigma}_z.
\eea
In this case, $[\hat{H}_s,\hat{S}] = 0$, and therefore the polaron shift operator commutes with the system's Hamiltonian. As a result, the system's  Hamiltonian is unchanged by the RCPT procedure.

(2) The standard spin-boson model is obtained when $\theta = \pi/2$,
\bea
\hat{H}_s^{eff}(\lambda,\theta=\pi/2) = \Delta e^{-\frac{2\lambda^2}{\Omega^2}} \hat{\sigma}_z.
\label{eq:effSB}
\eea
Here, the spin-splitting is {\it exponentially} suppressed due to the coupling to the environment, but
no new terms (processes) are generated in the system's Hamiltonian. This observation clearly points to the non-perturbative nature of the RCPT scheme.

(3) The intermediate angle $\theta = \pi/4$ leads to
\bea
\hat{H}_s^{eff}(\lambda,\theta=\pi/4) = \frac{\Delta}{2}(1+e^{-\frac{2\lambda^2}{\Omega^2}}) \hat{\sigma}_z + \frac{\Delta}{2} (1-e^{-\frac{2\lambda^2}{\Omega^2}}) \hat{\sigma}_x.
\nonumber\\
\eea
This intermediate case reveals the general features of strong coupling as predicted
by this technique: the qubit frequency is renormalized, 
similar to the standard spin-boson model, and a new tunneling term is generated. 


\begin{figure*}[htbp]
\centering
\includegraphics[width=2\columnwidth]{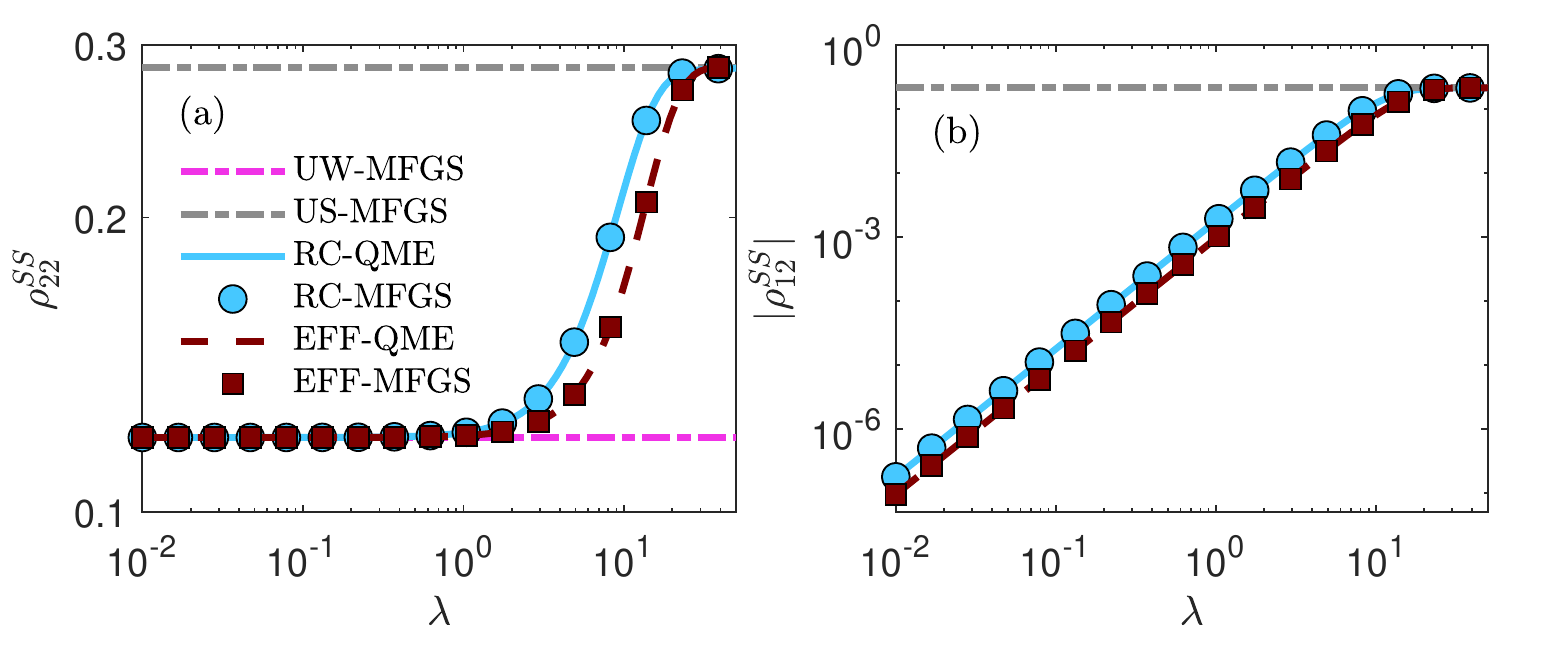}.
\caption{Thermalization in the generalized spin-boson model.
We present (a) the population and (b) the coherence in the eigenbasis of the Hamiltonian of the spin using 
two limitings cases of the mean force gibbs state: the ultraweak coupling limit Eq. (\ref{eq:UW-MFGS}), (UW-MFGS, magenta dashed-dotted), and the ultrastrong MFGS of Ref. \cite{Cresser} (US-MFGS light-grey dashed dotted). We further show the numerical RC-QME where the Redfield equation is solved after performing a reaction coordinate mapping (full), as well as its MGFS approximation, Eq. (\ref{eq:RCstate}) (cyan circles). We also show  the EFF-QME results where the Redfield QME is once again implemented after the RCPT method (dashed) as well as its MFGS approximation Eq. (\ref{eq:effMFGS}) (maroon squares).
 Parameters are $\Delta=1$, $\theta = \pi/4$ $\gamma=0.0071$, $\Omega = 20$, $\Lambda = 1000$, $T=\Delta/2$.}
\label{fig:Thermalization}
\end{figure*}

\subsection{Spectrum of the spin-boson model}

To showcase the predictive power of the RCPT method, we next compare the eigenenergy spectrum of the extended RC system, which is formally-exact and obtained from simulations, to the eigenenergies of the effective system Hamiltonian, which are inexact, but are given by analytical expressions. 
Our results are displayed in Figure \ref{fig:Spectrum} (a)-(c) for the three different angles, $\theta=\pi/2$, $\theta=0$, and $\theta=\pi/4$, respectively.

We diagonalize Eq. (\ref{eq:HSBeff}) and find the effective spin splitting for the three angles,
\bea
\Delta_{eff}(\lambda,\theta=0) &=& \Delta,
\label{eq:deltapi0}
\\
\Delta_{eff}(\lambda,\theta=\pi/2) &=& \Delta e^{-\frac{2\lambda^2}{\Omega^2}},
\label{eq:deltapi2}
\\
\Delta_{eff}(\lambda,\theta=\pi/4) &=& \frac{\Delta}{\sqrt{2}} \sqrt{ 1 + e^{-\frac{4\lambda^2}{\Omega^2}}}.
\label{eq:deltapi4}
\eea
In parallel, using the RC Hamiltonian, we study the energy spectrum of the extended system Hamiltonian, 
the first line of Eq. (\ref{eq:HRC}). 
We focus on the gap between the first excited state and the ground state. In Fig. \ref{fig:Spectrum} we show that this gap is perfectly reproduced by the energy differences of the effective Hamiltonian, as written in Eqs. (\ref{eq:deltapi0})-(\ref{eq:deltapi4}).
This agreement holds, surprisingly, even at very strong coupling with $\lambda>\Omega$.
We conclude that the RCPT technique provides an excellent approximation for the lowest energy levels of the system Hamiltonian, with strong-coupling effects absorbed in their definitions.
More broadly, as we show next, the method brings an intuition on the expected impact of strong-coupling effects in open system phenomena.



\subsection{Thermalization in the spin-boson model}

We now examine the long-time steady-state value of the system's density matrix as a function of the system-reservoir coupling parameter, $\lambda$.
Our main achievement here is the derivation of a closed-form analytic expression for the steady-state of the system, [Eq. (\ref{eq:effMFGS}) and Eq. 
(\ref{eq:rhoeffLambda}) in Appendix C], which is {\it exact} in both the weak and ultrastrong coupling limits. Moreover, it provides an excellent qualitative approximation to the steady state in the intermediate coupling regime.
 
In Figure \ref{fig:Thermalization} (a) and (b) we present the population of the excited state and the magnitude of the coherences of the spin, respectively, in the eigenbasis of the system Hamiltonian  for $\theta = \pi/4$ using $\Omega = 20$.
We present the elements of the density matrix using different methods: 

(i)  The ultra-weak MFGS (UW-MFGS) corresponds to the conventional Gibbs state,
\bea
    \rho_{UW-MFGS}^{SS} = \frac{1}{Z_{UW-MFGS}}e^{-\beta \hat{H}_s},
    \label{eq:UW-MFGS}
\eea
with $Z_{UW-MFGS}$ the partition function and
$\hat H_s$ the original system Hamiltonian, Eq. (\ref{eq:HSB}). 
It can be also shown that the Gibbs state is the
long time limit of the weak-coupling BMR-QME simulation \cite{Giacomo,Archak}.
In this limit, $\lambda$ dictates the {\it rate} to approach the steady state, but not its value, as we clearly see in  Fig. \ref{fig:Thermalization} (magenta dashed-dotted line). 

(ii) The RC-MFGS is defined as
\bea
 \rho_{RC-MFGS}^{SS} = \frac{1}{Z_{RC-MFGS}}\Tr_{RC}\left[e^{-\beta (\hat{H}_s + \Omega \hat{a}^{\dagger}\hat{a} + \lambda\hat{S}(\hat{a}^{\dagger} + \hat{a}))}\right],
 \nonumber\\
 \label{eq:RCstate}
 \eea
and it clearly has a nontrivial $\lambda$ dependence:
While  the {\it extended system}, which encompasses the RC, 
thermalizes to a conventional Gibbs state, the state of the spin itself, obtained after the RC is traced out, depends on $\lambda$. 
The RC-MFGS, Eq. (\ref{eq:RCstate}), is achieved numerically as the long-time solution of RC-QME simulations\cite{NazirPRA14}, and in
Figure \ref{fig:Thermalization} we present both calculations (cyan).
We clarify that the RC-QME value (cyan full line) is obtained by constructing the Redfield tensor and inverting it as in Eq. (\ref{eq:SSpop}).
In contrast, the RC-MFGS result (cyan circles)  is reached according to Eq. (\ref{eq:RCstate}) by constructing the extended system Hamiltonian (yet truncating the RC to include 11 levels, which is a sufficiently high number to represent the harmonic manifold of the RC), exponentiating the result, and tracing out the RC. These two approaches provide identical values. 

(iii)  The EFF-MFGS is
\bea
    \rho_{EFF-MFGS}^{SS} = \frac{1}{Z_{EFF-MFGS}}e^{-\beta \hat{H}_s^{eff}(\lambda)},
    \label{eq:effMFGS}
 \eea
with the effective system's Hamiltonian $\hat H_s^{eff}(\lambda)$ given by Eq. (\ref{eq:HSBeff}).
This state is tractable analytically, and it can be evaluated to give a closed-form expression, see Appendix C, culminating with Eq. (\ref{eq:rhoeffLambda}).
In Figure \ref{fig:Thermalization}, we present both the EFF-MFGS of Eq. (\ref{eq:effMFGS}) (maroon square) and steady state simulations based on the the EFF-QME method while using the effective Hamiltonian Eq. (\ref{eq:HSBeff})
(maroon, full lines). These two calculations agree and we find
that the steady state density matrix  depends on $\lambda$ in a non-trivial manner. 

Interestingly, the EFF-MFGS provides excellent qualitative results for {\it all coupling regimes}: It is exact in the asymptotically-weak coupling regime. It is also exact in the ultrastrong coupling limit (see 
Sec. \ref{sec:USEFFMFGS}). 
In between, it 
correctly reproduces the RC-MFGS
trends, albeit with some deviations in the position of the weak-to-strong crossover.


%
Concretely,  the steady state of the conventional spin-boson model ($\theta=\pi/2$) is diagonal with
 \bea
   \rho_{EFF-MFGS}^{SS}(\theta=\pi/2) \propto e^{-\beta \Delta_{eff}(\lambda)\sigma_z},
 \eea
where $\Delta_{eff}(\lambda)$ is given in Eq. (\ref{eq:deltapi2}). 
The steady state of the intermediate case, ($\theta = \pi/4$) presented in Figure \ref{fig:Thermalization} is nondiagonal, and thus maintains steady-state coherences, 
\bea
    \rho_{EFF-MFGS}^{SS}(\theta=\pi/4) \propto e^{-\frac{1}{2}\beta\Delta[(1+e^{-\frac{2\lambda^2}{\Omega^2}})\hat{\sigma}_z + (1-e^{-\frac{2\lambda^2}{\Omega^2}})\hat{\sigma}_x ]}.
    \nonumber\\
\eea
The proportionality constants in the above expressions are the reciprocal of the partition functions of the respective states, which can be  computed by a trace over the system.

(iv) The ultrastrong limit, US-MFGS, of
Ref. \citenum{Cresser}
is plotted as well in Fig. \ref{fig:Thermalization} (light grey dashed-dotted line). It is given below in Eq. (\ref{eq:USlimit}). Remarkably, the EFF-MFGS approaches this limit as $\lambda\to\infty$.
We discuss this limit in more detail in the following Section.

\subsection{Ultrastrong coupling limit of the generalized spin-boson model}
\label{sec:USEFFMFGS}

Focusing now on the ultrastrong coupling limit  with $\lambda\to \infty$, we obtain from the effective Hamiltonian - RCPT treatment [Eq. (\ref{eq:effMFGS})] the following steady state,
 \bea
\lim_{\lambda\to\infty}    \rho_{EFF-MFGS}^{SS} (\theta)\propto 
e^{-\frac{\beta\Delta}{2}\left[ (1 + \cos(2\theta))\hat\sigma_z + \sin(2\theta)\hat\sigma_x\right]}.
\label{eq:USeff}
 \eea
Thus, at very strong coupling, the conventional model ($\theta=\pi/2$) corresponds to the two levels being equally populated, with zero coherences. In contrast, when the coupling involves non-commuting operators using $\theta=\pi/4$, the RCPT method provides the steady state
 \bea
\lim_{\lambda\to\infty}    \rho_{EFF-MFGS}^{SS}(\theta=\pi/4) \propto e^{-\frac{\beta \Delta}{2}\left[\hat{\sigma}_z+\hat{\sigma}_x\right]}.
 \eea
Here, the equilibrium state possesses different populations from the standard spin-boson model, as well as  nonzero steady state coherences.

We now recall the ultrastrong limit of the MFGS (US-MFGS) derived in Ref. \citenum{Cresser} for the same model, 
\bea
\lim_{\lambda\to\infty}
&&\rho_{US-MFGS}^{SS} =
\nonumber\\
&&
\frac{1}{2}\left[  1-(\hat{\sigma}_x\sin(\theta)+\hat{\sigma}_z\cos(\theta)) \tanh(\beta\Delta \cos(\theta)) \right].
\label{eq:USlimit}
\nonumber\\
\eea
%
In Appendix C, we prove that the EFF-MFGS reduces to this expression in the $\lambda\to\infty$ limit.
It is significant to note that the elegant RCPT procedure produces results that {\it exactly} match the ultrastrong limit of Ref. \citenum{Cresser}.

\begin{figure}
    \centering
    \includegraphics[width=1\columnwidth]{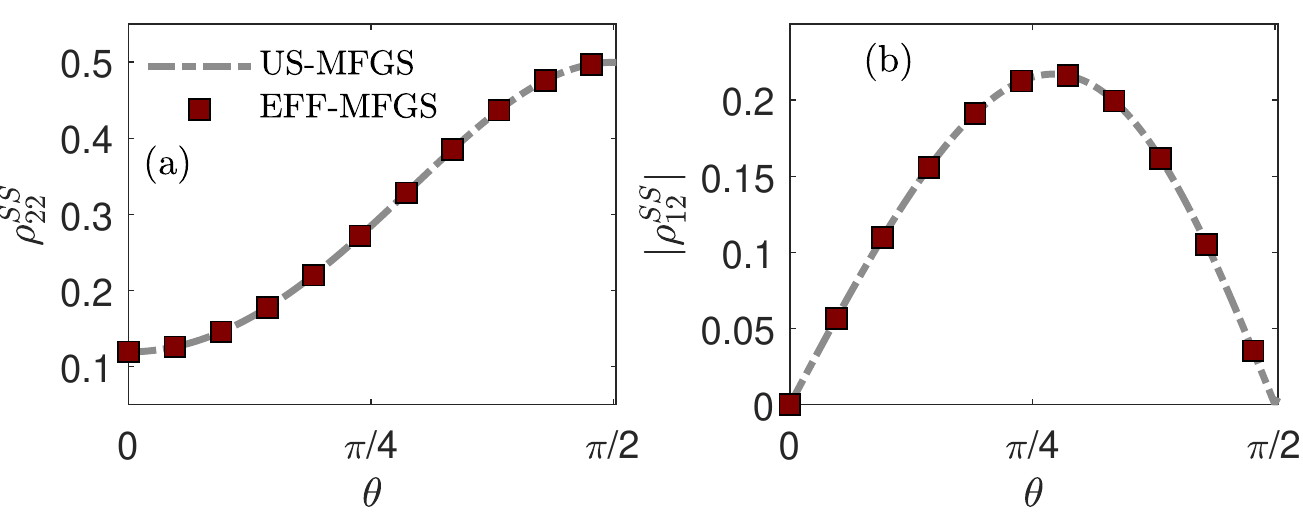}
\caption{The ultrastrong coupling limit of the generalized spin-boson model in steady state presenting the spin's (a) population and (b) coherence as a function of the angle $\theta$, which controls the noncommutativity of operators. We note on the exact agreement between the ultrastrong MFGS, Eq. (\ref{eq:USlimit}) (light-gray dashed-dotted), and our analytical results (maroon squares), calculated using Eq. (\ref{eq:USeff}).  Parameters are the same as in Figure \ref{fig:Thermalization}, with $\lambda \to \infty$.}
\label{fig:Ultrastrong}
\end{figure}

We further expand on the exact agreement between our RCPT approach and the ultrastrong limit of Ref. \citenum{Cresser} by showing in Figure \ref{fig:Ultrastrong} the steady state excited state populations and the magnitude of the coherences in the $\lambda \to \infty$ limit as a function of the  angle $\theta$. 
%
We briefly comment on this agreement, between EFF-MFGS and the 
US-MFGS, Eq. (\ref{eq:USlimit}): In Ref. \citenum{Cresser}, their result was derived by representing the Hamiltonian in the ``pointer basis", that is, the eigenbasis of the system's operator that is coupled to the bath. Projecting the effective system Hamiltonian Eq. (\ref{eq:HSeff}) to the pointer basis, we find that it is exactly equal to the pointer basis representation of the original, system Hamiltonian. Therefore, due to their pointer basis representations being the same, we should expect the same results for the two methods in the ultrastrong coupling limit. 

Figure \ref{fig:Ultrastrong} (a) shows an increase in the excited state population in the ultrastrong coupling limit with increasing $\theta$. As $\theta$ grows, the suppression of the spin-splitting becomes more substantial; in the limiting case of $\theta = \pi/2$, the ground and excited states are equally populated, since they become degenerate in the ultrastrong limit. For lower values of $\theta$ spin-splitting suppression is only one aspect of strong coupling, which explains why we stray from equally populated levels.
Furthermore, in Figure \ref{fig:Ultrastrong} (b) we observe that coherences are controlled by the angle $\theta$, with a maximum showing at
$\theta = \pi/4$. This can be traced back to the effective system Hamiltonian, Eq. (\ref{eq:HSBeff}), where a new contribution, a coupling-induced tunneling term, is maximized at this angle.

\subsection{Discussion and Extensions}

The principal far-reaching result of this section is the effective MFGS, Eq. (\ref{eq:effMFGS}) with its explicit form, Eq. (\ref{eq:rhoeffLambda}). This is a closed-form approximate analytic solution for the steady state density matrix that properly captures all coupling regimes, from the asymptotically weak to the ultra-strong limit. 
This example demonstrates that the effective model Hamiltonian, the outcome of the RCPT, provides an accurate description for the equilibrium state of a system coupled to a heat bath, covering the full range of coupling parameters, weak, intermediate and ultrastrong. 
The main advantage of the RCPT method is that the equilibrium state is readily obtained by performing the RCPT mapping, and there is no need to perform an actual open-system dynamics.

The EFF-MFGS can be calculated efficiently for other nontrivial models with steady state coherences and interactions.
The EFF-MFGS and the resulting partition function, allow us to obtain analytic expressions for thermodynamical observables (energy, heat capacity, entropy) in the strong-coupling limit.
For example, one could consider a fermionic analog of this study, a quantum dot model with an onsite Coulomb repulsion and strong coupling to the metals. Using the reaction coordinate method and developing a fermionic analog of the EFF-MFGS one may be able to evaluate electrical effects in the highly-correlated regime.

\section{Heat transport in the nonequilibrium spin-boson model}
\label{sec:qubitheat}

In this section we investigate the problem of quantum heat transport in the nonequilibrium spin-boson model, which provides a minimal setting to study heat transport at the nanoscale.
Such ideas of thermal transport were recently experimentally implemented using superconducting
quantum circuits \cite{Pekola1,Pekola2}.
The effective model provides an excellent analytical approximation to the quantum heat current, from weak to strong coupling, as was shown in Ref. \citenum{Nick2021}.

The nonequilibrium spin-boson model is identical to the generalized spin-boson model with $\theta = \pi/2$, 
except now the spin couples to two thermal reservoirs ($\alpha=L,R$) held at different temperatures; for a diagramatic representation, see Fig. \ref{fig:diagram-models}(b).
In this model, the two system operators that couple the spin to the different baths 
commute with each other, allowing for successive polaron transformations to be applied on the two RCs 
(extracted from each bath), with no conceptual complications.

\begin{figure}
    \centering
  \includegraphics[width=1\columnwidth]{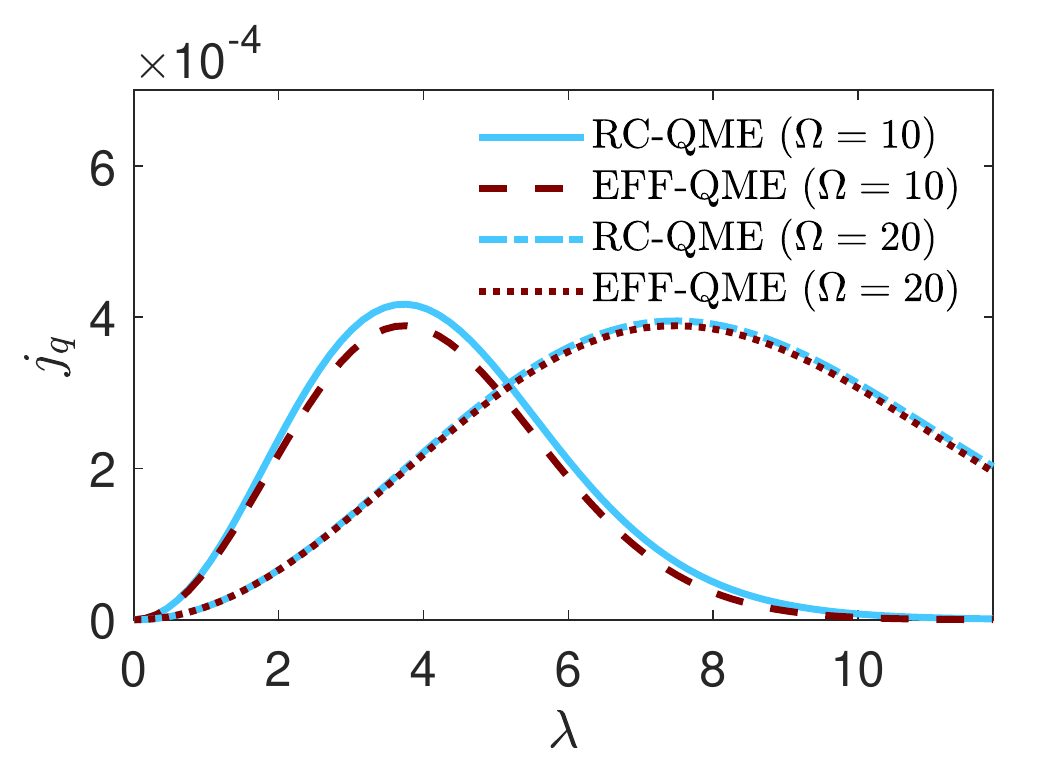}
\caption{
Quantum heat transport in the nonequilibrium spin-boson model.
We present the steady state heat current computed with the RC-QME method (cyan) and
the EFF-QME (maroon) 
at two different RC frequencies, $\Omega=10$ and $\Omega=20$. 
Parameters are $\Delta = 1$, $T_h = \Delta$, $T_c = \Delta/2$, $\gamma = 0.0071$, $\Lambda = 1000$. 
}
 \label{fig:Current}
\end{figure}

The Hamiltonian is given by
\bea
\hat{H} &= & \Delta \hat{\sigma}_z 
 \nonumber\\
&&+ \sum_{\alpha=\{L,R\},k} \nu_{\alpha,k} \left(\hat{c}_{\alpha,k}^{\dagger} + \frac{t_{\alpha,k}}{\nu_{\alpha,k}}\hat{\sigma}_x \right) \left(\hat{c}_{\alpha,k} + \frac{t_{\alpha,k}}{\nu_{\alpha,k}}\hat{\sigma}_x \right),\nonumber\\
\eea
The terms here are analogous to those in Eq. (\ref{eq:HSB}), 
except now there are two bosonic reservoirs, which are independent and maintained at different thermal states. 
We follow an identical procedure to Section \ref{sec:general}, but
 now we extract two RCs, one from each bath, and thus perform two polaron transformations, one for each RC. 
Since the two baths are coupled via the same system operator, 
we represent this transformation as a single polaron operator, 
\bea
\hat{U}_P = \hat{U}_{P,L}\hat{U}_{P,R} = e^{[\frac{\lambda_L}{\Omega_L}(\hat{a}_L^{\dagger} - \hat{a}_L) + \frac{\lambda_R}{\Omega_R}(\hat{a}_R^{\dagger} - \hat{a}_R) ]\hat{\sigma}_x }.
\eea
As a consequence of including an additional reservoir, the effective system Hamiltonian is
modified from Eq. (\ref{eq:effSB}). Namely, 
\bea
\hat{H}^{eff}_s = \Delta e^{-\sum_{\alpha=L,R}\frac{2\lambda^2_\alpha}{\Omega_\alpha^2}} \hat{\sigma}_z.
\label{eq:HeffSSB}
\eea
%
%
The total effective Hamiltonian of the model, Eq. (\ref{eq:Heff}) with Eq. (\ref{eq:HeffSSB}), is given by
\bea
&&\hat{H}^{eff}(\lambda) = 
\Delta e^{-\sum_{\alpha=L,R}\frac{2\lambda^2_\alpha}{\Omega_\alpha^2}} \hat{\sigma}_z
 \nonumber\\
&&+ \sum_{\alpha,k} \omega_{\alpha,k} \left(\hat{b}_{\alpha,k}^{\dagger} - \frac{2 \lambda_{\alpha} f_{\alpha,k}}{\Omega_{\alpha} \omega_{\alpha,k}}\hat{\sigma_x} 
\right) \left(\hat{b}_{\alpha,k} - \frac{2  \lambda_{\alpha} f_{\alpha,k}}{\Omega_{\alpha} \omega_{\alpha,k}}\hat{\sigma_x} \right).
\label{eq:Heff2}
\eea
Since in the effective model  the spin weakly couples to the heat bath, analytical expressions from the weak coupling limit were adopted
to  provide a closed-form expression for the heat current, capturing weak-to-strong coupling behavior \cite{Nick2021}.
We do not repeat these expressions here, but in  Fig. \ref{fig:Current} we present calculations of the heat current obtained from the RC-QME method and the EFF-QME technique.
Importantly, the two approaches are in an excellent agreement. 
This demonstrates that the effective treatment is appropriate for describing steady state properties,  even in the very-strong coupling regime. 
The main nontrivial observation from Fig. \ref{fig:Current}
is the turnover behavior of the heat current with coupling strength. This phenomenon  has been analyzed and demonstrated with powerful numerically-exact methods such as in Refs. \cite{Thoss08, Saito13, Tanimura16, Kilgour19},
 as well as with quantum master equation tools in the polaron frame, e.g.,
 \cite{PRL05,QME06,NIBA11,NIBA14,Cao1,Cao2,Cao3,Cao4,Wu20}. The RCPT method reproduces this nontrival behavior with minimal effort.
Fundamentally, we know that transport at weak coupling is sequential and resonant 
\cite{PRL05,NIBA11,QME06}. 
Inspecting Eq. (\ref{eq:Heff2}), we conclude that
 for large $\Omega$ transport is still sequential and resonant---yet with spin frequency that is monotonically quenched, revealing the origin of heat current suppression at strong coupling: When we increase $\lambda$ the current first increases due the enhancement in excitation and relaxation processes transferring  energy through the system. However, increasing $\lambda$ also suppresses the spin splitting, thus the quanta of energy transferred is being quenched.
 More details on heat transport in this model are given in Ref. \citenum{Nick2021}.

\section{Autonomous Quantum Absorption Refrigerator}
\label{sec:QAR}

An autonomous quantum absorption refrigerator extracts heat from a cold bath ($c$) and deposits it in a hot bath ($h$), assisted by heat input from a so-called work ($w$) reservoir, obeying $T_c<T_h<T_w$.
A canonical model for this machine is made of a quantum ``working fluid" with three energy states \cite{MitchisonReview,Hava2019,Junjie2021}, $|n\rangle$, $n=1,2,3$. For a schematic representation, see Fig. \ref{fig:diagram-models}(c). Transitions between the levels are achieved by absorbing or releasing heat to the different thermal reservoirs, with the following system operators, $\hat{S}_c = \ket{1}\bra{2} + h.c.$, $\hat{S}_w = \ket{2}\bra{3} + h.c.$, and $\hat{S}_h = \ket{1}\bra{3} + h.c.$. 
The total Hamiltonian of this model is
\bea
\hat{H} = \hat{H}_s + \sum_{\alpha=\{c,w,h\},k} \nu_{\alpha,k} \left(\hat c_{\alpha,k}^{\dagger} + \frac{t_{\alpha,k}}{\nu_{\alpha,k}}\hat{S}_\alpha\right) \left(\hat c_{\alpha,k} + \frac{t_{\alpha,k}}{\nu_{\alpha,k}}\hat{S}_\alpha\right).
\nonumber\\
\label{eq:HQAR}
\eea 
Here, $\hat c^{\dagger}_{\alpha,k}$ ($\hat c_{\alpha,k}$) are the bosonic creation (annihilation) operators to generate a quanta of frequency $\nu_{\alpha,k}$ in the $\alpha$th thermal bath; $t_{\alpha,k}$ are the system-bath coupling energies.
The system Hamiltonian is written in the energy basis as
\bea
\hat H_s=\sum_{n=1,2,3}\epsilon_{n}|n\rangle\langle n|.
\eea
For the system to act as a refrigerator, that is, extract heat from the cold bath and release it into the hot bath, one needs to tune the energy levels $\epsilon_{1,2,3}$.
Without loss of generality, below we use $\epsilon_1 = 0$, $\epsilon_2 = \Delta$ and $\epsilon_3 = 1$, and adjust $\Delta$ to achieve cooling.
While the cooling condition and the associated cooling current can be readily obtained assuming weak system-bath coupling \cite{Levy14,Luis14}, these calculations become nontrivial once we deviate from this assumption: 
In Ref. \citenum{QAR-Felix}, we used the RC-QME, a numerical tool, to locate the cooling window at strong coupling, revealing rich trends.

\begin{figure*}[htbp]
    \centering
  \includegraphics[width=1.8\columnwidth]{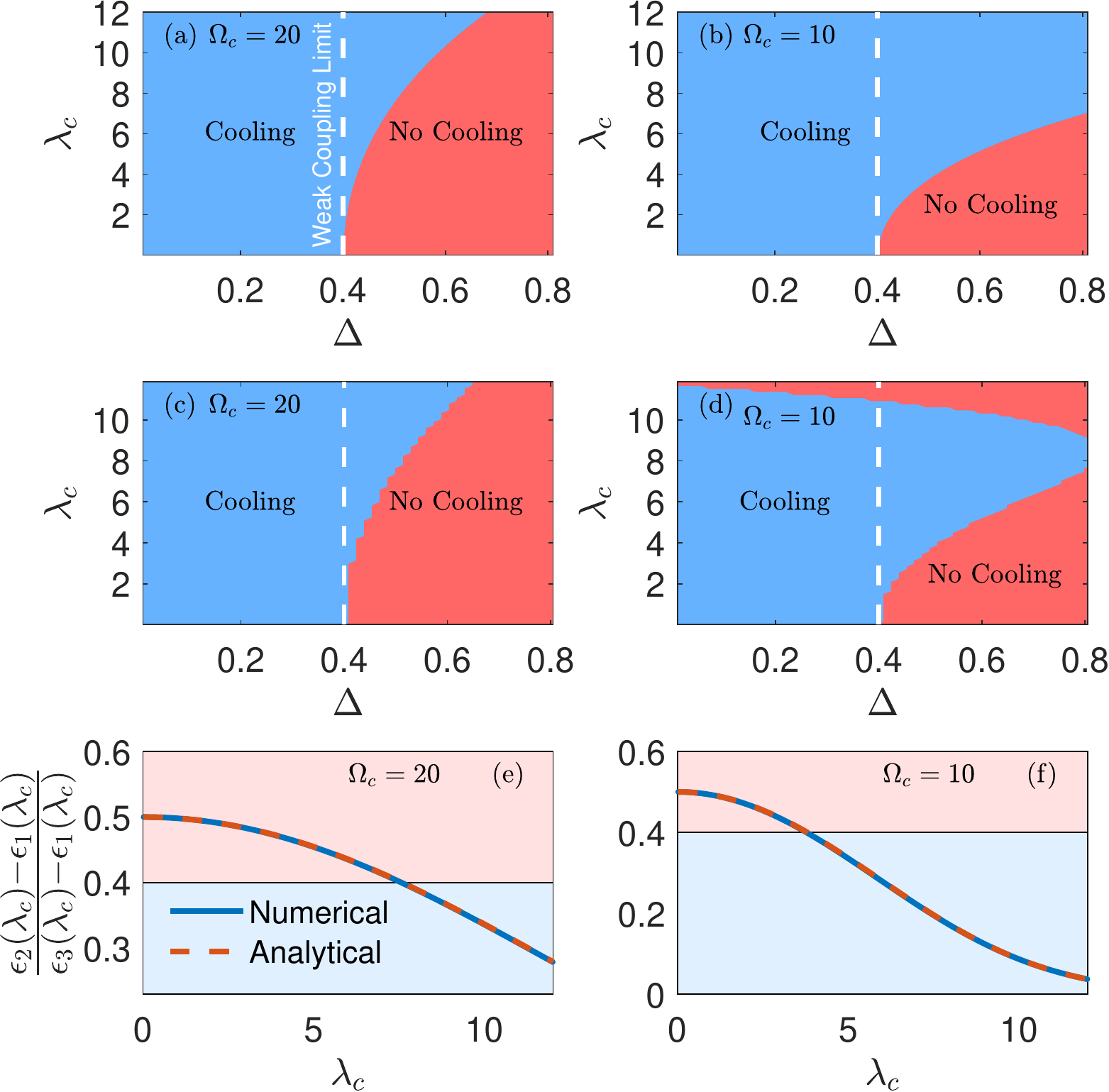}
\caption{Cooling window of the three-level autonomous quantum absorption refrigerator.
(a)-(b) The cooling window calculated {\it analytically} from the RCPT method using Eq. (\ref{eq:Window}) for $\Omega = 20$ (a) and $\Omega = 10$ (b). This is contrasted to panels (c) and (d) where the cooling window is calculated from the {\it numerical} RC-QME method.
The dashed line marks the boundary of the cooling window in the weak coupling limit, where cooling takes place for $0<\Delta<0.4$.
We used reservoir temperatures $T_c = 0.25$, $T_h = 0.5$, $T_w = 1.5$.
Panels (e) and (f) show a comparison of the eigenvalues calculated from the RCPT method and exact diagonalization of the RC Hamiltonian. Here, $\Delta = 0.5$. 
}
\label{fig:Window}
\end{figure*}
In what follows, we show that the RCPT approach can be used to provide an analytical expression for the cooling window --- assuming for simplicity that only the cold bath is strongly coupled to the three-level quantum system, while the other baths are weakly coupled to it.  
We thus extract a single reaction coordinate from the cold reservoir and apply the corresponding polaron transformation to the cold RC. This enables analytical expressions for the cooling condition to be obtained from the RCPT method, while not posing new challenges arising from the non-commuting polaron operators. 

We comment that we choose to extract the RC from the cold reservoir since it is at a lower temperature than the other baths, hence stronger correlations are expected to survive in this reservoir. The other two contacts are treated in the standard (BMR-QME) weak coupling fashion. 
The resulting Hamiltonian upon extracting a reaction coordinate from the cold reservoir is 
%
\bea
\hat{H}_{RC} &= &\hat{H}_s 
+ \sum_{\alpha=\{w,h\},k} \nu_{\alpha,k} \left(\hat c_{\alpha,k}^{\dagger} + \frac{t_{\alpha,k}}{\nu_{\alpha,k}}\hat{S}_\alpha\right) \left(\hat c_{\alpha,k} + \frac{t_{\alpha,k}}{\nu_{\alpha,k}}\hat{S}_\alpha\right)
\nonumber\\ 
&+&
\Omega_c \left(\hat{a}_c^{\dagger} +\frac{\lambda_c}{\Omega_c}\hat S_c \right)\left( \hat{a}_c +\frac{\lambda_c}{\Omega_c}\hat S_c \right)
%
\nonumber\\ 
&+& \sum_k \omega_{c,k} \left( \hat{b}_{c,k}^{\dagger} + \frac{f_{c,k}}{\omega_{c,k}} (\hat{a}_c^{\dagger}+\hat{a}_c)  \right)  \left( \hat{b}_{c,k} + \frac{f_{c,k}}{\omega_{c,k}} (\hat{a}_c^{\dagger}+\hat{a}_c)  \right).\nonumber\\
\label{eq:HRCQAR}
\eea
In this expression, the hot and work reservoirs are unchanged, compared to the initial model, Eq. (\ref{eq:HQAR}). The RC transformation acts exclusively on the cold reservoir. It extracts a collective coordinate from that bath of frequency $\Omega_c$, 
which couples to the system via $\lambda_c$. 
Next, we apply the polaron transformation, 
$\hat{U}_P = e^{\frac{\lambda_c}{\Omega_c}\hat{S}_c (\hat{a}_c^{\dagger} - \hat{a_c})}$, 
to (partially) decouple the cold RC from the three-level system. The resulting Hamiltonian is
\begin{widetext}
\bea
\hat{H}_{RC-P} = 
&&\hat{U}_P \hat{H}_s \hat{U}_P^{\dagger} + \sum_{\alpha=\{w,h\},k} \nu_{\alpha,k} \left(\hat c_{\alpha,k}^{\dagger} + \frac{t_{\alpha,k}}{\nu_{\alpha,k}} \hat{U}_P\hat{S}_\alpha\hat{U}_P^{\dagger}\right) \left(\hat c_{\alpha,k} + \frac{t_{\alpha,k}}{\nu_{\alpha,k}}\hat{U}_P\hat{S}_\alpha\hat{U}_P^{\dagger}\right)
\\ \nonumber
&+& 
\Omega_c \hat a_c^{\dagger}\hat a_c
+ \sum_k \omega_{c,k} \left(\hat{b}_{c,k}^{\dagger} + \frac{f_{c,k}}{\omega_{c,k}} (\hat{a}^{\dagger}_c + \hat{a}_c - \frac{2\lambda_c}{\Omega_c}\hat{S}_c)  \right) \left(\hat{b}_{c,k} + \frac{f_{c,k}}{\omega_{c,k}} (\hat{a}^{\dagger}_c + \hat{a}_c - \frac{2\lambda_c}{\Omega_c}\hat{S}_c)  \right).
\eea 
%
Focusing on the subspace with zero excitations in the RC, we arrive at our effective description of the 3-level QAR Hamiltonian (ignoring constant shift terms),
%
\bea
\hat{H}^{eff}(\lambda) &= &\hat{H}_s + \frac{\Delta}{2}\left(e^{-\frac{2\lambda_c^2}{\Omega_c^2}}-1\right)\hat{Q} + \sum_{\alpha=\{w,h\},k} \nu_{\alpha,k} \left(\hat{c}_{\alpha,k}^{\dagger} + \frac{t_{\alpha,k}}{\nu_{\alpha,k}}e^{-\frac{\lambda^2_c}{2\Omega_c^2}}\hat{S}_\alpha\right) \left(\hat{c}_{\alpha,k} + \frac{t_{\alpha,k}}{\nu_{\alpha,k}}e^{-\frac{\lambda^2_c}{2\Omega_c^2}} \hat{S}_\alpha\right)
\nonumber\\
&+& \sum_k \omega_{c,k} \left( \hat{b}_{c,k}^{\dagger} - \frac{2\lambda_c f_{c,k}}{\Omega_c \omega_{c,k}}\hat{S}_c \right) \left( \hat{b}_{c,k} - \frac{2 \lambda_c f_{c,k}}{\Omega_c \omega_{c,k}}\hat{S}_c \right).
\label{eq:HeffQAR}
\eea
\end{widetext}
In this expression, the operator $\hat{Q} = -\ket{1}\bra{1} + \ket{2}\bra{2}$ arises from the action of the polaron transformation on the system Hamiltonian and it represents a shift of the first two energy levels of the QAR.


Inspecting Eq. (\ref{eq:HeffQAR}),
the overall effect of strong system-bath coupling at the cold contact as observed from the RCPT treatment is nontrivial:
(i) The energy difference between the lowest two energy levels (those coupled to the cold bath) is suppressed, 
$\Delta\to \Delta e^{-\frac{2\lambda_c^2}{\Omega_c^2}}$.
This effect is similar to the suppression of the spin spacing in the spin-boson model, Eq. (\ref{eq:deltapi2}).
(ii) Transitions in the system that are induced by the hot and work baths are suppressed by the cold bath. This effect is highly nontrivial. 

The cooling condition specifies regimes in which the system can act as a refrigerator and extract heat from  the cold environment. In the weak coupling limit and using the Born-Markov Redfield quantum master equation, the cooling condition is
\cite{Levy14,Luis14}
\bea
\frac{\epsilon_2 - \epsilon_1}{\epsilon_3 - \epsilon_1} 
\leq \frac{\beta_h - \beta_w}{\beta_c - \beta_w}.
\label{eq:coolcond}
\eea
The effect of strong coupling is to {\it dress} the QAR parameters. In particular, the energy levels of the QAR, $\epsilon_n$, gain a dependence on $\lambda_c$.
The renormalized energy levels are [see Eq. (\ref{eq:HeffQAR})],
\bea
\epsilon_1(\lambda_c) &=& \frac{\Delta}{2}\left(1 - e^{-\frac{2\lambda^2_c}{\Omega_c^2}}\right),
\\ 
\epsilon_2(\lambda_c) &=& \frac{\Delta}{2}\left(1+e^{-\frac{2\lambda^2_c}{\Omega_c^2}}\right),
\\ 
\epsilon_3(\lambda_c) &=& 1.
\eea
%
The cooling condition Eq. (\ref{eq:coolcond}) was derived for the original Hamiltonian (\ref{eq:HQAR}) under the weak coupling condition. It thus holds for the effective Hamiltonian
(\ref{eq:HeffQAR}) since it has the same form, only with renormalized parameters, and with weak coupling restored between the  effective system and the bath.
We thus write down the cooling condition in the strong coupling regime using the renormalized levels, 
\bea
\frac{\epsilon_2(\lambda) - \epsilon_1(\lambda)}{\epsilon_3(\lambda) - \epsilon_1(\lambda)} = \frac{\Delta e^{-\frac{2\lambda_c^2}{\Omega_c^2}}}{1 - \frac{\Delta}{2}\left(1 - e^{-\frac{2\lambda_c^2}{\Omega_c^2}}\right)} \leq \frac{\beta_h - \beta_w}{\beta_c - \beta_w}.
\label{eq:Window}
\eea
The gap between the lowest two energy levels is suppressed faster with $\lambda_c$ than the total gap. As a result, at large $\Delta$, where cooling was impossible at weak coupling, we observe cooling once we reach the strong system-reservoir coupling regime.
This effect is seen in Figure \ref{fig:Window}: The cooling window calculated using Eq. (\ref{eq:Window}) is displayed in Figure \ref{fig:Window}(a) and (b). It is compared to the cooling window predicted by the weak coupling limit (to the left of the dashed line at $\Delta=0.4$). 
This analytic result is also compared to 
 numerical simulations with the RC-QME method,
see Figure \ref{fig:Window}(c) and (d).

In these figures, the blue region corresponds to areas where cooling is allowed ($j_c > 0$) whereas red regions identifies the no cooling regime ($j_c \leq 0$). We find that the effective treatment agrees well with complete simulations for large RC frequency, $\Omega_c = 20$, while for smaller $\Omega_c = 10$, the agreement is not as good, particularly at large $\lambda_c$ values. This is to be expected since the RCPT analytical approach relies on $\Omega_c$ being the largest energy scale in the problem, and deviations  are expected once $\lambda_c \approx \Omega_c$. 

We comment that  deviations between the two approaches are not attributed to problems in capturing the eigenspectrum of the QAR at strong coupling. Figure \ref{fig:Window} (e) and (f) show the LHS of the cooling inequality, and we compare the analytical expression Eq. (\ref{eq:Window}) with the effective energies
to the value computed by taking the three lowest eigenvalues of the Hamiltonian  (\ref{eq:HRCQAR}). We observe perfect agreement even at large $\lambda_c$. 
This correspondence reveals that the RCPT method fails to capture the cooling window at small $\Omega$ due to transitions missing in the method, the result of the energy truncation involved. For example, leakage effects, with heat flowing directly from the work to the cold bath are missing in the effective Hamiltonian \cite{QAR-Felix}.

Equation (\ref{eq:Window}) demonstrates the remarkable predictive power of the RCPT method. 
Since strong coupling effects are now embedded in the energy levels of the system, a wealth of results describing performance bounds on weakly-coupled systems can be effortlessly extended to the strong coupling regime. 



%


\section{Phonon-assisted thermoelectric engines}
\label{sec:thermoE}

In this Section, we explore another nontrivial application of the RCPT technique 
to obtain a deeper understanding of phonon-assisted electron transport and thermoelectric generation in quantum dot setups. In this model, the RC and the subsequent polaron transformation are applied to a bosonic (phonon) reservoir 
that is strongly coupled to the system's electronic degrees of freedom (quantum dots). These quantum dots are assumed to weakly hybridize with voltage-biased and temperature-biased fermionic environments (metals) responsible for both charge and energy currents flowing in the junction. A schematic representation of the model is given in Fig. \ref{fig:diagram-models}(d).

As we show in this Section, using the RCPT method on the phonon-assisted charge transport model we gain three outcomes: (i) We bypass expensive simulations while treating strong coupling effects nonperturbatively. 
(ii) We analytically distill impacts of strong couplings from the renormalization of parameters in the effective Hamiltonian.
(iii) We achieve closed-form expressions for transport characteristics, here focusing on the efficiency of a thermoelectric power generator. 
As for physical observables, the RCPT method provided excellent predictions not only for the averaged charge current, but also for its fluctuations, as well as for the energy current.


\subsection{Model and the derivation of the Effective Hamiltonian}

The literature includes many theoretical proposals for phonon-assisted quantum-dot based thermoelectric generators, for example, Refs. \citenum{JianHuaImry,junjie2019,Aharony2010,Thoss2011,Esposito2013,McConnel22}.
In Refs. \citenum{Simine12,BijayTE1,BijayTE2}, for instance, 
phonon-assisted conduction and thermoelectric generation
were analyzed in double quantum dot devices. In those studies, however, the hybridization of the dots to the metal electrodes was assumed strong, but the electronic states of the quantum dots only perturbatively coupled to phonons; computationally extensive simulations in Ref. \citenum{SiminePI} explored nonperturbative electron-phonon coupling effects.

In the present study and following Ref. \citenum{McConnel22}, 
we assume that the coupling of the quantum dots to the metal electrodes is weak and can be handled in a perturbative manner by a second-order BMR-QME. The coupling of the quantum dot to a phonon bath is however strong, and this interaction, which will be treated with the RCPT method is essential for facilitating charge transport. 

The Hamiltonian of the double quantum dot is written in the 
${\ket{G},\ket{L},\ket{R},\ket{D}}$ basis, 
which corresponds to the states with neither dots being occupied,  the left dot only occupied, the right dot only occupied,  and both dots  occupied, respectively.
In this basis, the total Hamiltonian is represented as
\begin{widetext}
\bea
\hat{H} = && \epsilon_L \ket{L}\bra{L} + \epsilon_R \ket{R}\bra{R} + (\epsilon_L + \epsilon_R + U) \ket{D}\bra{D}  
 + \sum_k \epsilon_{k,L} \hat{c}_{k,L}^{\dagger} \hat{c}_{k,L} + \sum_k \epsilon_{k,R} \hat{c}_{k,R}^{\dagger} \hat{c}_{k,R}
\nonumber\\
&+& \sum_k \left[ (\ket{R}\bra{D} - \ket{G}\bra{L})h_{k,L} \hat{c}_{k,L}^{\dagger} + h.c. \right ] + \sum_k \left [ (\ket{L}\bra{D} + \ket{G}\bra{R})h_{k,R} \hat{c}_{k,R}^{\dagger} + h.c. \right] 
\nonumber\\
&+& \sum_q \nu_q \left( \hat{d}_q^{\dagger} + \frac{t_q}{\nu_q} (\ket{L}\bra{R} + h.c.) \right) 
\left( \hat{d}_q + \frac{t_q}{\nu_q} (\ket{L}\bra{R} + h.c.) \right).
\label{eq:Hephonon}
\eea
\end{widetext}
For more details on this model, see Ref. \citenum{McConnel22}.
In the above expression, $\epsilon_{L,R}$ are the energies of the left and right quantum dots, and  
$U$ is the Coulomb interaction energy when both quantum dots are occupied.  The fermionic reservoirs 
are coupled to the dots with a coupling strength $h_{k,L/R}$;
here the creation (annihilation) operators $\hat{c}_{k,\alpha}^{\dagger}$ ($\hat{c}_{k,\alpha}$)
create (annihilate) an electron in  the fermionic lead $\alpha=L,R$ with energy $\epsilon_{k,\alpha}$. We assume a linear dispersion relation for the electronic energy with a wideband  of constant density of states.
The last line in Eq. (\ref{eq:Hephonon}) describes electron tunneling between the two dots --- assisted by a phonon bath. 
The phononic degrees of freedom are described by  creation (annihilation)  operators  $\hat{d}_{q}^{\dagger}$ ($\hat{d}_{q}$).
Here, $q$ identifies a normal mode with frequency $\nu_q$
 coupled to electronic transitions between the dots with the coupling energy $t_q$. 
 

Following Ref. \citenum{McConnel22}, we introduce a compact notation for the system operators  on the double quantum dot Hilbert space,
$\hat{A}_1 = -\ket{G}\bra{L} + \ket{R}\bra{D}$, 
$\hat{A}_2 = -\ket{L}\bra{G} + \ket{D}\bra{R}$, 
$\hat{A}_3 = \ket{G}\bra{R} + \ket{L}\bra{D}$, 
$\hat{A}_4 = \ket{R}\bra{G} + \ket{D}\bra{L}$, 
$\hat{S} = \ket{L}\bra{R} + \ket{R}\ket{L}$, 
$\hat{L} = \ket{L}\bra{L}$, 
$\hat{R} = \ket{R}\bra{R}$, and 
$\hat{D} = \ket{D}\bra{D}$. 
We allow phonons to strongly couple to electrons, thus we
perform a reaction coordinate transformation on the phononic degrees of freedom to extract a collective phonon coordinate and add it to the system's Hamiltonian. The dot-metal hybridization is assumed weak (though in principle one can also perform the RC mapping on the electronic energies).

After the polaron transform and the truncation of the RC mode,
we arrive at our effective Hamiltonian,  exhibiting strong electron-phonon coupling through renormalized parameters and new coupling terms.
After neglecting constant terms we obtain
%
\begin{widetext}
\bea
\nonumber
\hat{H}^{eff}(\lambda) &= &\left[\epsilon_L \cosh(\frac{\lambda^2}{\Omega^2}) 
+ \epsilon_R \sinh(\frac{\lambda^2}{\Omega^2})  \right]e^{-\frac{\lambda^2}{\Omega^2}}\hat{L} 
+ \left[\epsilon_R \cosh(\frac{\lambda^2}{\Omega^2}) 
+ \epsilon_L \sinh(\frac{\lambda^2}{\Omega^2})  \right]e^{-\frac{\lambda^2}{\Omega^2}}\hat{R} 
+ (\epsilon_L + \epsilon_R + U)\hat{D}
\nonumber \\
&+& \sum_q \omega_q \left( \hat{b}_q^{\dagger} - \frac{2 \lambda f_q}{\Omega \omega_q} \hat{S}\right) 
\left( \hat{b}_q - \frac{2 \lambda f_q}{\Omega \omega_q} \hat{S}\right) + \sum_k \left[ \hat{A}_1 h_{k,L} e^{-\frac{\lambda^2}{2\Omega^2}} \hat{c}_{k,L}^{\dagger} + \hat{A}_2 h^*_{k,L}e^{-\frac{\lambda^2}{2\Omega^2}} \hat{c}_{k,L} \right ] 
\nonumber\\ 
&+& \sum_k \left [ \hat{A}_3 h_{k,R}e^{-\frac{\lambda^2}{2\Omega^2}} \hat{c}_{k,R}^{\dagger} + \hat{A}_4 h^*_{k,R}e^{-\frac{\lambda^2}{2\Omega^2}} \hat{c}_{k,R} \right] + \sum_k \epsilon_{k,L} \hat{c}_{k,L}^{\dagger} \hat{c}_{k,L} + \sum_k \epsilon_{k,R} \hat{c}_{k,R}^{\dagger} \hat{c}_{k,R}.
\label{eq:HephononEff}
\eea
\end{widetext}
Here, $\Omega$  and $\lambda$ are parameters of the spectral density function of the phonon bath, describing the central frequency of the bath and its coupling energy to the electronic system, see Eq. (\ref{eq:Brownian}). However, after the RCPT procedure, these bath parameters are imprinted into the model Hamiltonian itself. Furthermore, since we assume that the spectral density function is narrow, and that $\lambda<\Omega$, the residual phonon bath only weakly couples to the system, as in Eq. (\ref{eq:Jeff}). Intermediate steps in the calculation are presented in Appendix D.
 
Inspecting the Hamiltonian (\ref{eq:HephononEff}), and in comparison to the original expression Eq. (\ref{eq:Hephonon}), the effects of the RCPT mapping can be summarized as follows:
(i) The energy levels of the electronic dots are renormalized by the coupling to phonons such that they approach equal values at strong coupling.
(ii) The coupling of the phonon bath to the dots is dressed (weakened) by the factor $\lambda/\Omega$. 
(iii) Electron tunneling from the metals to the dots is exponentially suppressed.
\begin{figure}[htbp]
    \centering
    \includegraphics[width=1\columnwidth]{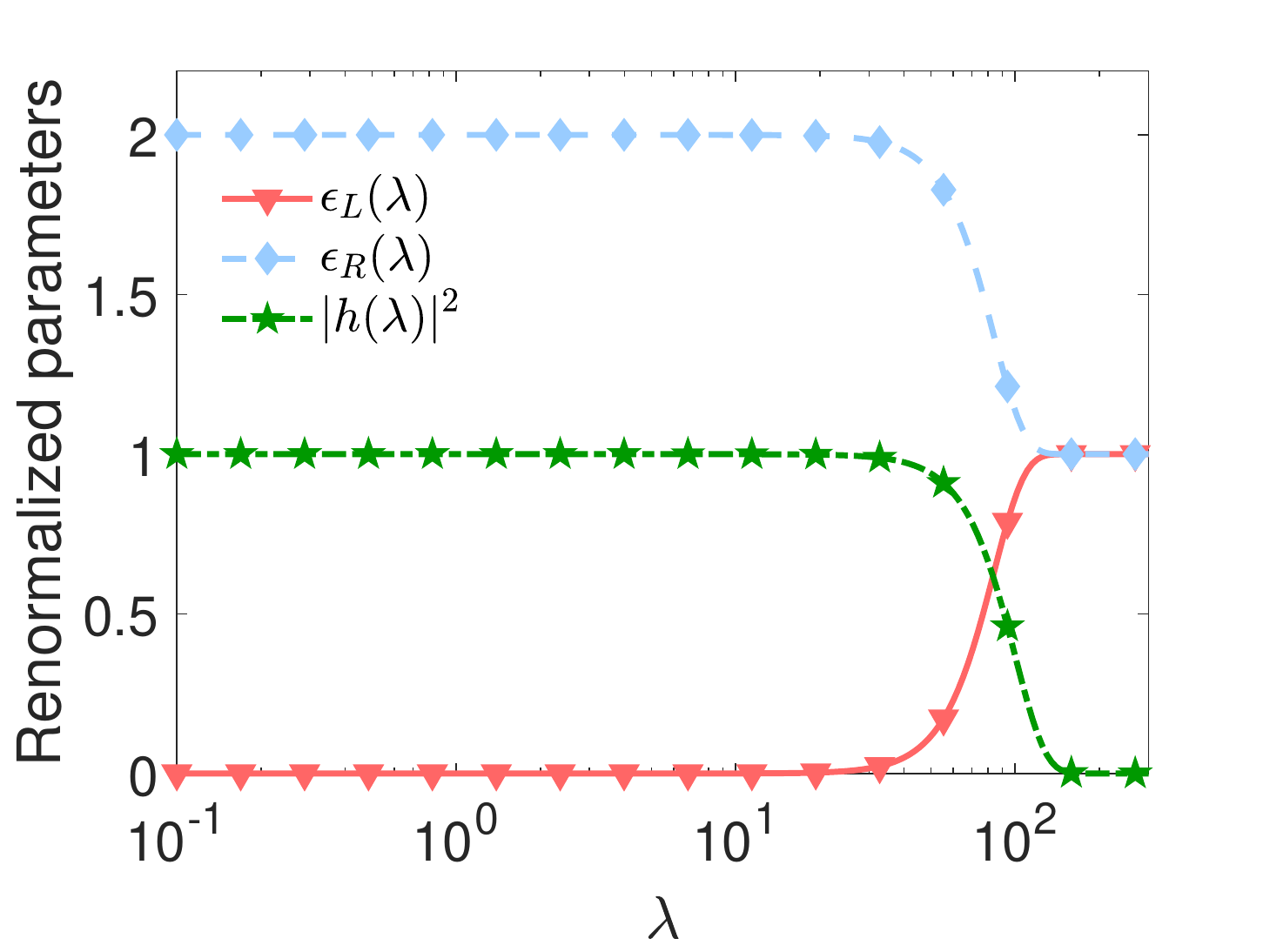}
\caption{Effective coupling-dressed parameters in the phonon-assisted quantum dot thermoelectric generator.
We display the energy of the left dot (triangle), right dot (diamond) and the magnitude squared of the coupling energy between the dots and the fermionic baths (star), as calculated from Eqs. (\ref{eq: ELeff})-(\ref{eq:heff}).  
Parameters (without dressing) are $\epsilon_R=2$, $\epsilon_L=0$, $|h|^2 = 1$, $\Omega=100$.}
\label{fig:EffparamsthermoE}
\end{figure}


\begin{figure*}[htbp]
 \centering
 \includegraphics[width=2\columnwidth]{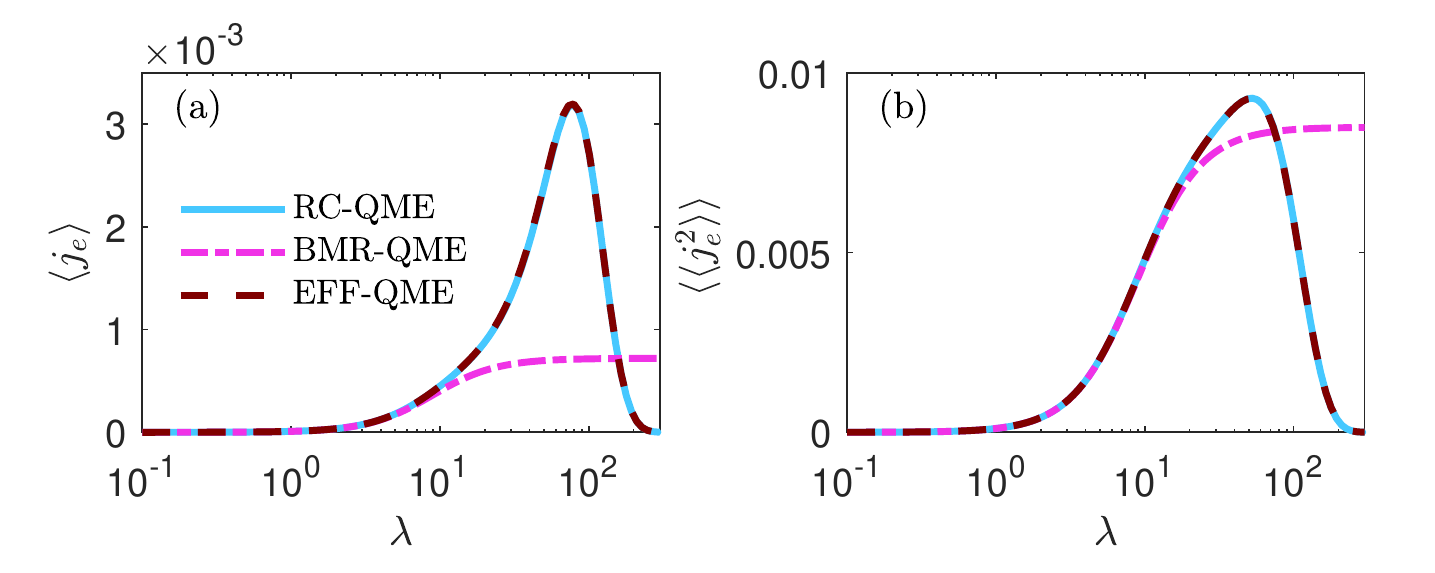}
\label{fig:Thermoelectric}
\caption{Phonon-assisted quantum dot thermoelectric power generator.
(a) Mean  steady state charge current flowing left to right (positive), and (b) the current fluctuations. 
Parameters are $\epsilon_R=2$, $\epsilon_L=0$,
$T_L=10 $, $T_R=1$, $\Omega=100$, $\mu_L=-0.3$ $\mu_R=-0.2$, $V= \mu_R-\mu_L=0.1$,
$T_{ph}=1$, $2\pi\gamma\Omega=100$, and metal-dot hybridization energies
$\Gamma_L=\Gamma_R=0.1$, 
all in units of $T_R$. 
}
\label{fig:thermoE}
\end{figure*}

\begin{figure*}[htbp]
    \centering
  \includegraphics[width=2\columnwidth]{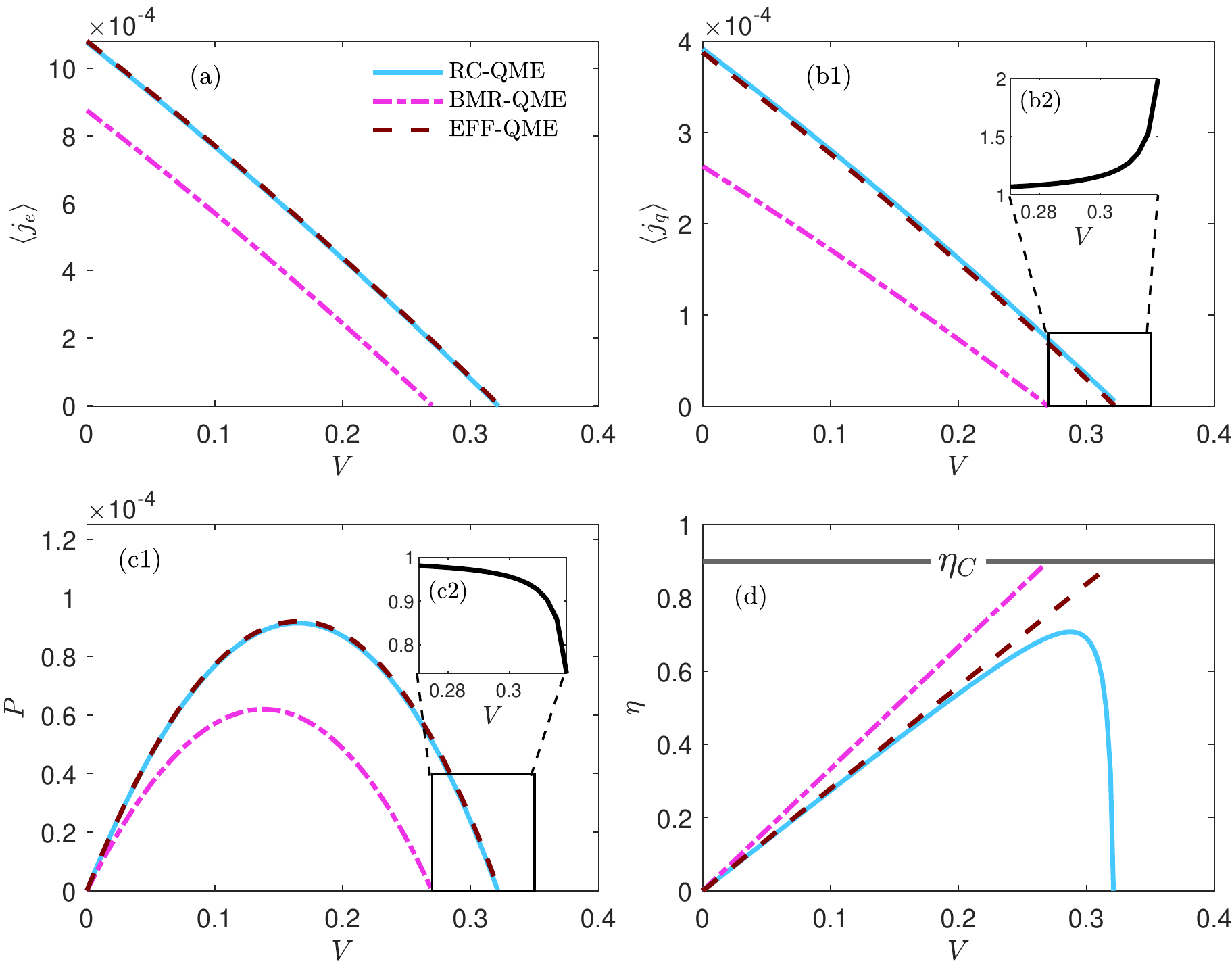}
    \label{fig:ThermoelectricVoltage}
\caption{Phonon-assisted double quantum dot thermoelectric power generator.
(a) Mean charge current from the left lead (positive) towards the system and (b) mean heat current from the left reservoir. 
(c) Power output of the generator, which is given by the charge current times voltage, and (d) the thermoelectric efficiency, compared to the Carnot bound $\eta_C$. The insets (b2) and (c2) present the ratio of currents (RC-QME result over EFF-QME) at the vicinity of the stopping voltage.
Parameters are the same as in Figure \ref{fig:thermoE}, with $\lambda = 17.3$ and $\mu_L = -0.3$, while $\mu_R$ is varied.
}
\label{fig:thermoEVoltage}
\end{figure*}


To expound the impact of strong system-bath couplings, we define the renormalized energy parameters
\bea
&&\epsilon_L(\lambda) =  \left[\epsilon_L \cosh(\frac{\lambda^2}{\Omega^2}) 
+ \epsilon_R \sinh(\frac{\lambda^2}{\Omega^2})  \right]e^{-\frac{\lambda^2}{\Omega^2}},
\label{eq: ELeff}
\\
&&\epsilon_R(\lambda) = \left[\epsilon_R \cosh(\frac{\lambda^2}{\Omega^2}) 
+ \epsilon_L \sinh(\frac{\lambda^2}{\Omega^2})  \right]e^{-\frac{\lambda^2}{\Omega^2}},
\label{eq:EReff}
\\
&&h_{k,L}(\lambda) = h_{k,L}e^{-\frac{\lambda^2}{2\Omega^2}}; \,\,\,\,\,\,\,\
h_{k,R}(\lambda) = h_{k,R}e^{-\frac{\lambda^2}{2\Omega^2}},
\label{eq:heff}
\eea
corresponding to the phonon-dressed quantum dots energies,
$\epsilon_{L,R}(\lambda)$ and the phonon-dressed metal-dot hybridizations, $h_{k,L}(\lambda)$ and $h_{k,R}(\lambda)$.

In Fig. \ref{fig:EffparamsthermoE} we show these renormalized parameters, which are strongly affected by the electron-phonon coupling when $\lambda$ approaches $\Omega$.
For a system configured with $\epsilon_R = 2$ and $\epsilon_L = 0$, we again observe level renormalization as a staple of strong coupling in this model. Here, the quantum dot energies approach their average value at strong coupling. Furthermore, we also observe a suppression of the dot-metal hybridization  as the coupling of electrons to the phonon bath is increased, which is notable, since the RC mapping did not involve the fermionic reservoirs. This is indeed a polaronic effect, with the electrons being slowed down due to polaron formation on the dots. 
As we show below with benchmarking, the RCPT method performs extremely well in this model. It quantitatively captures the significant features of the model even as $\lambda$ becomes comparable to $\Omega$, a regime that is not guaranteed to be properly described by the RCPT.

\subsection{Charge current and its noise}

We now turn to study charge transport in the model.
The quantum dot system is coupled to two metal electrodes and a phonon bath, and one could use this setup to investigate numerous aspects of quantum transport such as the behavior of the charge current and its fluctuations as a function of voltage and electron-phonon couplings, with all baths maintained at the same temperature. The system can be also tuned to act as a thermoelectic power generator when applying a temperature difference counteracting the voltage bias.
To study this function, we follow Ref. \citenum{McConnel22} and investigate the same setup. The left metal is set hot and the right side is cold, $T_L>T_R$. However, the chemical potentials of the electrodes are tuned with the opposite polarity, $\mu_L<\mu_R$.
As for the temperature of the phonon bath $T_{ph}$, we set it here to be equal to $T_R$, but one could imagine other situations as described in Ref. \citenum{McConnel22}. 
The metal molecule hybridization is defined as  
\bea
\Gamma_{L}(\epsilon)=2\pi\sum_{k}|h_{k,L}|^2\delta(\epsilon-\epsilon_{k,L}),
\eea
and a similar expression is used to define $\Gamma_R(\epsilon)$. We assume that these parameters are energy independent, and we work in the weak metal-dot coupling limit such that  $\Gamma_{L,R}\ll T_{L,R}$. 
As for the phonon bath, it is described by a Brownian spectral function with the peak frequency at $\Omega$, Eq. (\ref{eq:Brownian}).
After the mapping, the residual bath couples weakly to the quantum dots, with an Ohmic spectral function.

In Fig. \ref{fig:thermoE} we display the mean charge current and its fluctuations as a function of the electron-phonon coupling strength, $\lambda$. 
We calculate the charge current using Eq. (\ref{eq:curre}), presenting it here with the brackets, $\langle j_e\rangle $, to emphasize that this is the mean current;
using a full-counting statistics approach, we also calculate the current noise, denoted here by
$\langle \langle j_e^2\rangle\rangle = \langle j_e^2\rangle- \langle j_e\rangle^2$. 
Technical details on how to calculate currents and noise in the model are given in Ref. \citenum{McConnel22} and we do not repeat them here.

We present results using three methods: BMR-QME, which is valid at weak electron-phonon coupling only, RC-QME, a numerical tool simulating transport [based on Eq. (\ref{eq:HephononRC})], which is expected to hold even for large $\lambda$, and the EFF-QME method using the effective Hamiltonian (\ref{eq:HephononEff}), then simulating current with the BMR-QME method.
Focusing in Fig. \ref{fig:thermoE} on trends as a function of electron-phonon coupling strength, we note the perfect agreement between the latter two techniques, showcasing the excellent performance of the RCPT method compared to full simulations.

The RCPT method is not only remarkably computationally efficient (as we do not need to pay any computational cost for working in the strong coupling limit), but furthermore it clarifies the origins of (i) the significant enhancement of the current at intermediate electron-phonon coupling compared to the weak coupling limit, and (ii) the complete suppression of charge current at the ultrastrong coupling limit, as we discuss next.

At weak electron-phonon coupling, the current trivially grows with $\lambda$ due the increasing coupling between the dots and the phonon bath (as in the weak coupling scheme) assisting transport.  At the intermediate regime (here around $\lambda=50$) the current shoots up, contrasting with the behavior at weak coupling. The reason for this strong enhancement of the current is made clear when looking at the effective Hamiltonian (\ref{eq:HephononEff}): As we increase the electron-phonon coupling the energy levels of the quantum dots approach degeneracy, reaching their mean value $(\epsilon_L+\epsilon_R)/2$ in the ultrastrong coupling limit.
Evening the energy levels --- closing their gap --- is beneficial for charge transport.
However, at the same time, the metal-quantum dot tunneling elements $|h(\lambda)|^2$ are exponentially suppressed with $\lambda$. In the polaron picture this effect is well known: Unlike the bare electron, an electron dressed by lattice vibrations stabilizes and it requires  the ``reorganization energy" to hop. The combination of these effects lead to the turnover behavior of the current, and its eventual exponential suppression with $\lambda$.
It is significant to note that besides the mean current, its fluctuations are also excellently captured by the RCPT method, similarly showing a corresponding turnover behavior.

\subsection{Thermoelectric efficiency at strong coupling: Simulations and analytic results} 

Using the RCPT formalism we next simulate the charge $\langle j_e\rangle$ and energy currents  $\langle j_u\rangle$ arriving from the hot metal, as well as the associated heat current $\langle j_q\rangle = \langle j_u\rangle-\mu_L \langle j_e\rangle $. Combining these currents, we assess the efficiency of the thermoelectric generator, defined as
\bea
    \eta \equiv \frac{P}{\langle j_q \rangle},
    \label{eq:eff}
\eea    
with the power extracted  $P\equiv\langle j_e\rangle (\mu_R-\mu_L)$.
The efficiency is bounded by the Carnot limit,
$\eta_C=1-\frac{T_c}{T_h}$ with $T_{c,h}$ as the temperatures of the cold and hot baths.
Nontrivial questions concern how the thermoelectric efficiency depends on voltage, and how it is modified by the electron-phonon coupling energy.


In Fig. \ref{fig:thermoEVoltage} we 
look at the dependence of the mean charge current, mean heat current, and power output on the applied voltage bias between the right and left leads ($V = \mu_R-\mu_L$). Here, $\mu_R$ is varied while $\mu_L$ is kept constant. We immediately note the excellent agreement between the RC-QME and EFF-QME methods in Fig. \ref{fig:thermoEVoltage} (a)-(c). 
Using the data for currents, in Fig. \ref{fig:thermoEVoltage} (d) we plot the thermoelectric efficiency based on Eq. (\ref{eq:eff}). 
We observe the following: The BMR-QME method predicts that the efficiency grows linearly with voltage reaching the Carnot bound. Indeed, according to a weak coupling master equation theory, the efficiency of a thermoelectric generator is given by $\eta = \frac{\mu_R-\mu_L}{\epsilon_L - \mu_L}$. This reflects the 
tight coupling limit between the charge and heat currents, resulting in their cancellation from the expression for efficiency.
Obviously, since the electron-phonon coupling strength is large, the BMR-QME prediction is provided here as a reference point only.
Contrasting the characteristic linear trend of weak coupling, RC-QME simulations show that at finite electron-phonon coupling, the system cannot reach the Carnot efficiency, and the efficiency drastically drops to zero as we reach the stopping voltage.
Interestingly, the RCPT method with EFF-QME simulations provide accurate results for small-intermediate voltage biases, but it fails to capture the suppression of efficiency at higher voltage when approaching the stopping voltage. In other words, the EFF-QME method predicts that the efficiency can still reach the Carnot bound, only
with a different slope---due to the renormalization of parameters. This observation is consistent with the nature of the EFF-QME method.
It deploys a weak-coupling theory on an effective Hamiltonian, thus allowing reaching the Carnot bound:
In the ultrastrong coupling limit and based on Eq. (\ref{eq:heff}) the efficiency is given by
$\eta_{US}= \frac{\mu_R-\mu_L}{(\epsilon_L+\epsilon_R)/2 - \mu_L}$, distinct from the weak-coupling prediction 
by the value in the denominator. 

It is intriguing to note that while both charge and heat currents are seemingly excellently reproduced by the EFF-QME method compared to simulations with RC-QME, see Fig. \ref{fig:thermoEVoltage}(a)-(c), the corresponding thermoelectric efficiencies display marked differences.
These deviations can be understood from the inset plots, Fig. \ref{fig:thermoEVoltage} (b2) and (c2), where we note small deviations in both the heat current and power:
According to Fig. \ref{fig:thermoEVoltage} (b2) the heat current of the RC-QME approaches zero at a slightly higher voltage than the EFF-QME. Conversely, in Fig. \ref{fig:thermoEVoltage} (c2) we find that the power output predicted by the RC-QME tends towards zero at a slightly lower voltage as compared with the EFF-QME. 
This effect can also be understood as a difference in stopping voltages. The weak-coupling BMR-QME  predicts a stopping voltage of  $V_s = (\epsilon_L - \mu_L)(\beta_R - \beta_L)/\beta_R$ 
with $\beta_{L,R}=1/T_{L,R}$ \cite{McConnel22}. However, at strong coupling the stopping voltage increases (in the ultrastrong limit, $\epsilon_L\to \frac{\epsilon_L+\epsilon_R}{2} $). This is because the energy level of the left quantum dot increases with $\lambda$ due to strong coupling renormalization, see Fig. \ref{fig:EffparamsthermoE}. 

\subsection{Discussion}
The RCPT method shows excellent predictive power when describing the charge and heat currents and their fluctuations, even at very strong electron-phonon couplings and beyond its rigorous regime of applicability, extending to $\lambda\approx \Omega$.
The measure of thermoelectric efficiency, in contrast, 
is sensitive to small deviations. Since the RCPT method still captures only the tight coupling (proportionality) of the currents, its predictions for the efficiency miss the turnover behavior near the stopping voltage.
There are many model variants of phonon-assisted charge transport, including the celebrated Anderson-Holstein model. Employing the RCPT framework on canonical models, further including strong hybridization of the dots to the leads, could expose the rich physics of dissipative, correlated, nonequilibrium fermionic systems.


\section{Dissipative spin chains}
\label{sec:chain}

Quantum spin models serve a central role in our understanding of quantum many-body systems, specifically universal aspects of quantum phase transitions in magnetic systems. More recently, {\it dissipative} spin chains have been studied in e.g., Refs. \cite{SchallerReview,spinc1,spinc2,spinc3,Prosen2011,Landi2014,Pereira2018}, motivated by applications in
quantum information processing and spintronics, as well as
real-world experiments simulating spin lattices with cold atoms \cite{Bloch}.

We show here that the RCPT method can be readily used to study the properties of dissipative spin chains, namely their spin polarization and heat transport behavior.
We present the theory on a 1-dimensional $N$-site Heisenberg model under a magnetic field. In simulations, we exemplify the theory on a two qubit system coupled via a general XYZ Ising interaction. Our main result is that
due to the impact of strong dissipation, the XX model approaches the Ising model Hamiltonian at strong bath coupling. Thus, dissipation can mask distinct features of spin chain models. 

The Hamiltonian of a dissipative Heisenberg chain with $N$ sites is written as
\bea
\hat{H} &=&
\sum_{\alpha=1}^{N}\Delta_\alpha \hat{\sigma}_z^{\alpha} 
+ \sum_{i\in\{x,y,z\}}\sum_{\alpha=1}^{N-1} J_i \hat{\sigma}_i^{\alpha} \hat{\sigma}_i^{\alpha+1}  
\nonumber\\
&+& \sum_{\alpha=1}^N\sum_k \nu_{\alpha,k} \left( \hat{c}_{\alpha,k}^{\dagger} + \frac{t_{\alpha,k}}{\nu_{\alpha,k}}\hat{\sigma}_x^{\alpha} \right) \left( \hat{c}_{\alpha,k} + \frac{t_{\alpha,k}}{\nu_{\alpha,k}}\hat{\sigma}_x^{\alpha} \right). 
\label{eq:Hspinchain}
\eea
In this expression, $\Delta_{\alpha}$ represents the spin splitting of the $\alpha$th qubit. The qubits are coupled to each other with strength $J_{x,y,z}$ along the different directions.
The qubits are also each coupled to a local bosonic reservoir with modes of frequency $\nu_{\alpha,k}$ at strength $t_{\alpha,k}$. We assume as before that the spectral density functions of these baths are of Brownian form,
Eq. (\ref{eq:Brownian}), with $\Omega_{\alpha}$ and $\lambda_{\alpha}$ the centre of the Brownian functions
and the respective qubit-bath coupling strength.
Proceeding via the RCPT protocol as outlined in Sec. \ref{sec:general}, we arrive at the effective Hamiltonian, 
\begin{widetext}
\bea
\hat{H}^{eff}(\lambda_{1},\lambda_2,...,\lambda_N) = && \sum_{\alpha=1}^N\Delta_{\alpha} e^{-\frac{2\lambda_{\alpha}^2}{\Omega_{\alpha}^2}}  \hat{\sigma}_z^{\alpha} 
+ \sum_{\alpha=1}^{N-1} \left[J_x \hat{\sigma}_x^{\alpha} \hat{\sigma}_x^{\alpha+1} %
+ J_y e^{-\frac{2\lambda_{\alpha}^2}{\Omega_{\alpha}^2}}
e^{-\frac{2\lambda_{\alpha+1}^2}{\Omega_{\alpha+1}^2}} \hat{\sigma}_y^{\alpha} \hat{\sigma}_y^{\alpha+1}  
+
J_z e^{-\frac{2\lambda_{\alpha}^2}{\Omega_{\alpha}^2}} e^{-\frac{2\lambda_{\alpha+1}^2}{\Omega_{\alpha+1}^2}} \hat{\sigma}_z^{\alpha} \hat{\sigma}_z^{\alpha+1} 
\right]
 \nonumber\\
&+& \sum_{\alpha=1}^N\sum_k \omega_{\alpha,k} \left(\hat{b}_{\alpha,k}^{\dagger} - \frac{2\lambda_{\alpha} f_{\alpha,k}}{\Omega_{\alpha} \omega_{\alpha,k}} \hat{\sigma}_x^{\alpha}\right) \left(\hat{b}_{\alpha,k} - \frac{2\lambda_{\alpha} f_{\alpha,k}}{\Omega_{\alpha} \omega_{\alpha,k}} \hat{\sigma}_x^{\alpha}\right).
\label{eq:H2Qeff}
\eea
\end{widetext}
The effective Hamiltonian depends on all $\lambda_{\alpha}$ and $\Omega_{\alpha}$, though we highlight the dependence on the former. For details on the intermediate steps in the RCPT method, see Appendix E.
The residual coupling of the qubits to their baths is now weak, as explained in Section \ref{sec:general}. 

Inspecting the Hamiltonian, Eq. (\ref{eq:H2Qeff}), we note that the suppression of the spin splitting and the interaction parameters $J_y$ and $J_z$ due to the coupling to the baths results in the dissipative XX-model approaching the Ising model in the strong coupling regime.
Hence, we discover that models that behave distinctively when isolated from their surroundings become more and more similar as the system-bath interaction is increased. 
%

In what follows, we study the equilibrium and transport properties of the Hamiltonian in Eq. (\ref{eq:H2Qeff}) considering only two spins, denoted by $L$ and $R$, and for two special cases: (i) A transverse-field Ising type interaction where $J_z = J_y = 0$, and $J_x = J$. (ii) An XX type interaction with $J_z = 0$, and $J_y = J_x = J$. 
In particular, two-qubit models can be used as components for thermal energy transport, with each qubit coupled to a heat bath at a different temperature \cite{Geraldine1,Geraldine2}.
For a schematic representation, see Fig. \ref{fig:diagram-models}(e).

\begin{figure}
    \centering
  \includegraphics[width=1.0\columnwidth]{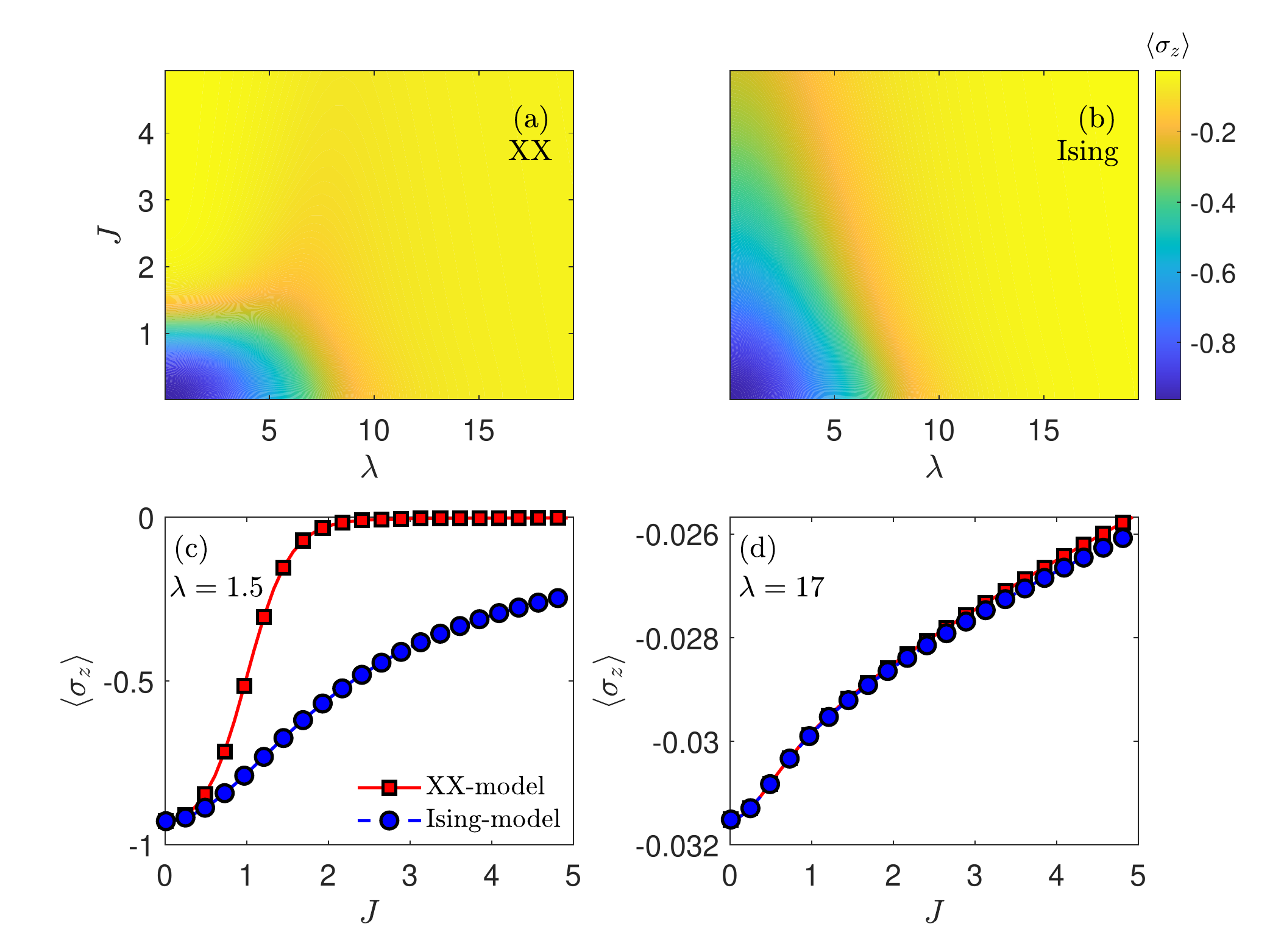}
\caption{Equilibrium magnetization in a two-qubit model, plotted as a function of the interaction strength $J$ and the coupling parameter $\lambda$ using the RC-QME method for both the XX-type (a) and Ising type (b) interactions. Panels (c) and (d) show cuts of the contour at weak and strong $\lambda$, respectively. Parameters are $\Delta_L = \Delta_R = 1$, $T_L = T_R =0.5$, and $\Omega = 10$.}
    \label{fig:Magnetization-XX}
\end{figure}

\begin{figure}
    \centering
  \includegraphics[width=1.0\columnwidth]{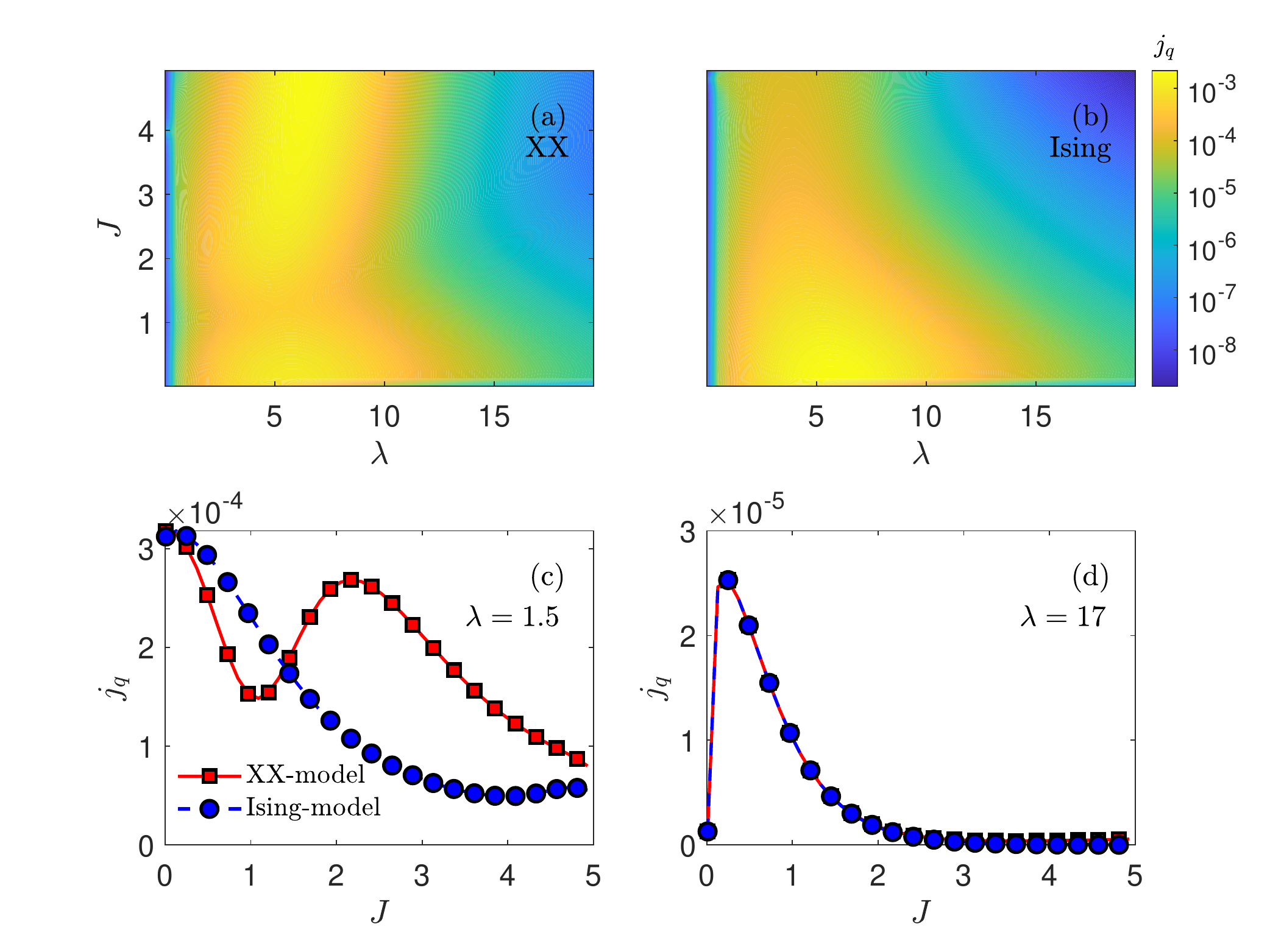}
\caption{Steady-state heat current through a two-qubit system, plotted as a function of the interaction strength $J$ and the coupling parameter $\lambda$ using the RC-QME method for both the XX-type (a) and Ising type (b) interactions. Panels (c) and (d) show cuts of the contour at weak and strong $\lambda$, respectively. Parameters are $\Delta_L = \Delta_R = 1$, $T_L = 0.5$, $T_R = 1$, $\Omega = 10$.}
    \label{fig:Current-XX}
\end{figure}

Considering the first line of the Hamiltonian, Eq. (\ref{eq:H2Qeff}), in the large coupling limit, all but the term proportional to $J_x$ will be exponentially suppressed. Therefore:
(i) In the strong coupling limit, the eigenstates of the effective system Hamiltonian coincide with the eigenstates of $\hat{\sigma}_x^L$ and $\hat{\sigma}_x^R$. This  implies that the XX-type model at strong coupling reduces to a description of the Ising model, with zero qubit splitting. 
(ii) 
When the qubits are coupled to heat baths at different temperatures, heat current can flow between the baths through the qubits. However, the heat current will be suppressed at strong system-bath coupling since (in the ultrastrong limit) the effective system Hamiltonian commutes with the total Hamiltonian, implying that energy cannot flow in the system.
These predictions are arrived at by simply inspecting terms in the effective Hamiltonian, Eq. (\ref{eq:H2Qeff}). In Figures  \ref{fig:Magnetization-XX} and \ref{fig:Current-XX} we test these predictions using the numerical RC-QME method.

Focusing first on an  equilibrium setting with the two baths set at the same temperature, in Fig.  \ref{fig:Magnetization-XX} we present the spin magnetization as a function of the spin-spin interaction $J$ and the system-bath coupling strength $\lambda$. 
We find that while at weak coupling the XX and the Ising models behave differently as a function of the exchange interaction $J$, both models follow similar trends as one increases the couplings to the heat baths.
Specifically, at weak coupling the XX model shows a transition from a polarized state to an unpolarized state at $J \approx 1.5$; the Ising model in contrast slowly looses polarization with increasing $J$. At strong dissipation, in contrast, the two models similarly preserve small polarization irrespective of the coupling strength, which is expected given the bath-induced quenching of the spin splitting. 

In Figure \ref{fig:Current-XX}, we turn to a nonequilibrium steady state situation with $T_L\neq T_R$ and 
 present the current of the XX model (a)  and the Ising model (b). The relevant quadrant is in the strong $\lambda$, weak $J$ limit, where current predicted by the XX and Ising models coincide. In this regime, since $J$ is weak relative to $\Omega$, our RCPT effective treatment is relevant, and our prediction of the two models coinciding in their behavior and leading to suppressed currents are verified in simulations.
Deviations in the currents supported by the two models are apparent in the upper right quadrant, which corresponds to the large $J$, large $\lambda$ limit. Here, since $J$ becomes comparable to $\Omega$, the  RCPT framework starts to break down, and our predictions of the XX model mapping into the Ising model are not as accurate. 

Altogether, the effective Hamiltonian treatment is a powerful new tool towards studying dissipative spin chains. Besides allowing feasible numerical simulations, the strength of the method lies in it directly building effective Hamiltonians that expose the impact of dissipation on model parameters, thus on the expected equilibrium phases and transport properties of these paradigmatic systems.

\section{Conclusion}
\label{sec:conclusion}

We introduced the reaction-coordinate polaron-transformation framework, an analytical-numerical tool for tackling open quantum system problems at strong system-bath coupling. This approach is applicable to a broad range of open quantum systems.
While computationally-expensive techniques have been developed in recent years to handle strong-coupling effects, including HEOM, chain mapping, tensor network and path integral approaches, the RCPT method stands out with it offering fundamental understanding as to the different impacts of strong couplings, as well as a route for highly-economic and reasonably accurate numerical simulations. 
While the method was introduced and exercised here for systems linearly coupled to the displacements of bosonic-harmonic environments, the approach can be extended to treat strong coupling effects between fermionic degrees of freedom.

The essence of our RCPT procedure is that 
 strong system-bath interactions were absorbed and embedded into a (modified) system Hamiltonian, which itself became weakly-coupled to its surroundings, thus allowing economical simulations and analytical derivations. 
The procedure involved performing a reaction coordinate mapping to extract the prominent degrees of freedom from the bath, applying next a polaron transformation to partially decouple the RC from the system, and finally truncating the reaction coordinate. These three steps resulted in an effective model Hamiltonian with strong system-bath coupling built into the system. 

We employed the RCPT method and studied central questions in quantum thermalization, quantum transport and quantum thermodynamics. Focusing on the steady state regime, the RCPT method allowed us to predict and rationalize trends, derive closed-form expressions quantifying the performance of many-body quantum thermal machines, and perform economic simulations. We exemplified the capacity of the RCPT method with five paradigmatic problems:

(i) We investigated the topic of {\it quantum thermalization} using the generalized spin-boson model in Sec. \ref{sec:qubit}. Our main result via the RCPT procedure was the derivation of a closed-form expression for the thermal equilibrium state of the system. This expression  is {\it exact}  in both the weak and the ultrastrong coupling limits, and it further provides qualitatively correct results in the intermediate regime.

(ii) Quantum heat transport was investigated using the generalized spin-boson model in Sec. \ref{sec:qubitheat}.
The RCPT approach provided the characteristic turnover of the heat current with system-bath coupling energy.

(iii) The impact of strong coupling on the cooling performance of continuous quantum absorption refrigerators was analyzed in Sec. \ref{sec:QAR}. Here, the RCPT method allowed us to derive analytical expressions for the cooling window, exposing the role of strong coupling.

(iv) The problem of phonon-assisted electron transport was studied in Sec. \ref{sec:thermoE}, with a focus on the performance of thermoelectric power generators. The RCPT method provided accurate predictions not only of the charge current, but also its fluctuations, revealing a turnover behavior when increasing the coupling to phonons. The method further allows us to write a closed-form expression for the efficiency of the phonon-assisted power generator, valid from linear response to the far-from-equilibrium region (yet, as expected, missing the correct behavior near the stopping voltage).

(v) Dissipative quantum chains were analyzed in Sec. \ref{sec:chain}.
The RCPT method revealed the confluence of different spin-chain models once dissipation was enhanced.

These five canonical models embody many-body interactions, include strong system-bath coupling effects, and encompass rich physics from linear response to the far-from-equilibrium regime. The powerful RCPT method elegantly captured their equilibrium physics and transport characteristics with little effort. 

%
%
%

We focused in this study on the steady state behavior of quantum thermal machines. In future work we plan to look at how effective models deal with transient dynamics. In this context, complications arise as the RCPT method may neglect non-Markovian effects due to the truncation of the reaction coordinate. As such, in order to accurately study quantum dynamics, a method that can capture such features is required.
Another potential avenue for the RCPT method is the application of iterative mappings, making use, for example, of numerical spectral density functions for the RC transformation. This would allow studies with spectral density functions that are richer, beyond the Brownian form.
More broadly, we envision the development and application of RCPT-inspired mapping methods to understand and simulate light-matter systems, time-dependent driven materials, and interacting fermionic models.



\begin{acknowledgements}
DS acknowledges support from an NSERC Discovery Grant and the Canada Research Chair program.
The work of NAS was supported by Ontario Graduate Scholarship (OGS).
We acknowledge fruitful discussions with Janet Anders.
\end{acknowledgements}


\begin{widetext}
\renewcommand{\theequation}{A\arabic{equation}}
\setcounter{equation}{0}  
\setcounter{section}{0} 
\section*{Appendix A: Calculation of the effective system Hamiltonian}
\label{app:1}
In this appendix, we simplify the expression obtained in Sec. \ref{sec:general} of the main text for the system portion within the total effective Hamiltonian. Our starting point is the subspace of the polaron-dressed system Hamiltonian, (Eq. (\ref{eq:HSeff0}) in the main text),
\bea
\hat{H}_{s}^{eff}(\lambda) = \bra{0}e^{\frac{\lambda}{\Omega} (\hat{a}^{\dagger} - \hat{a})\hat{S}} \hat{H}_{s} e^{-\frac{\lambda}{\Omega} (\hat{a}^{\dagger} - \hat{a})\hat{S}} \ket{0}.  \eea
The polaron transformation operator has a similar mathematical structure to the displacement operator, $D(\alpha) = e^{\alpha \hat{a}^{\dagger} - \alpha \hat{a}}$, where in our situation, the parameter $\alpha$ is in fact an operator living in the Hilbert space of the system,
$\alpha \equiv \frac{\lambda}{\Omega}\hat{S}$. We note, that here, $\alpha$ is hermitian. 
We use the following properties of the displacement operator: $D(-\alpha) = D^{\dagger}(\alpha)$, and $D(\alpha)\ket{0} = \ket{\alpha}$ which implies that $D^{\dagger}(\alpha)\ket{0} = \ket{-\alpha}$ to write the effective system Hamiltonian as
\bea
    \hat{H}_{s}^{eff}(\lambda) &=& \bra{-\alpha}\hat{H}_s\ket{-\alpha}.
\eea
Next, we comment that coherent states can be represented by the  eigenstates of the harmonic oscillator $|n\rangle$ as
$\ket{\alpha}=e^{-\frac{|\alpha|^2}{2}}
\sum_{n=0}^{\infty} \frac{\alpha^n}{\sqrt{n!}}|n\rangle$.
In the models examined here, all the elements in $\alpha$ are real thus we ignore the absolute value symbol.
Combining these facts, we compute the effective system Hamiltonian,
\bea
\hat{H}_{s}^{eff}(\lambda) &=& e^{-\frac{\alpha^2}{2}}\sum_{n,m} \frac{(-1)^n}{\sqrt{n!}} \bra{n}\alpha^n \hat{H}_s \alpha^m \ket{m} \frac{(-1)^m}{\sqrt{m!}} e^{-\frac{\alpha^2}{2}}
\nonumber\\
&=& e^{-\frac{\alpha^2}{2}}\sum_{n} \frac{(-1)^{2n} }{n!} \alpha^n \hat{H}_s \alpha^n e^{-\frac{\alpha^2}{2}},
\nonumber\\
&=& e^{-\frac{\lambda^2}{2\Omega^2}\hat{S}^2} \left( \sum_{n} \frac{\lambda^{2n} }{\Omega^{2n} n!} \hat{S}^n \hat{H}_s \hat{S}^n \right ) e^{-\frac{\lambda^2}{2\Omega^2}\hat{S}^2}.
\eea
This is our final expression: In the second line, we made use of the fact that $\hat{H}_s$ and $\hat{S}$ are operators that act on the system Hilbert space only. Therefore, the partial matrix element resolves simply to a Kronecker product in the $m$ and $n$ states.

\renewcommand{\theequation}{B\arabic{equation}}
\setcounter{equation}{0}  
\setcounter{section}{0} 
\section*{Appendix B: Effective Hamiltonian Higher Order Contributions}
\label{app:2}

In this appendix, we explain how to systematically extend the RCPT method and build
higher order contributions to the effective Hamiltonian of the system, Eq. (\ref{eq:Heff1}). This is done by including higher order excitations to the RC manifold.
For example, if we include two levels to the RC, $|0\rangle$ and $|1\rangle$, the  effective Hamiltonian becomes a $2\times 2$ matrix,
\bea
    \hat{H}_s^{eff,[2]} &=& \bra{0} e^{\frac{\lambda}{\Omega} (\hat{a}^{\dagger} - \hat{a})\hat{S}} \hat{H}_{s} e^{-\frac{\lambda}{\Omega} (\hat{a}^{\dagger} - \hat{a})\hat{S}} \ket{0} \ket{0}\bra{0}
    \nonumber\\
    &+&
     \bra{1} e^{\frac{\lambda}{\Omega} (\hat{a}^{\dagger} - \hat{a})\hat{S}} \hat{H}_{s} e^{-\frac{\lambda}{\Omega} (\hat{a}^{\dagger} - \hat{a})\hat{S}} \ket{1} \ket{1}\bra{1}
      \nonumber\\
    &+&
 \bra{0} e^{\frac{\lambda}{\Omega} (\hat{a}^{\dagger} - \hat{a})\hat{S}} \hat{H}_{s} e^{-\frac{\lambda}{\Omega} (\hat{a}^{\dagger} - \hat{a})\hat{S}} \ket{1} \ket{0}\bra{1}
      \nonumber\\
    &+&
   \bra{1} e^{\frac{\lambda}{\Omega} (\hat{a}^{\dagger} - \hat{a})\hat{S}} \hat{H}_{s} e^{-\frac{\lambda}{\Omega} (\hat{a}^{\dagger} - \hat{a})\hat{S}} \ket{0} \ket{1}\bra{0}.
\eea
In the main text, we limited the occupation number of the RC to zero, assuming $\Omega\gg T$. Here, we compute as an example the matrix element between the $k$th and $p$th levels. 
Such extensions to higher occupations of the RC should allow for a more complete description of the RCPT technique and provide corrections for better numerical accuracy. Therefore, as an extension of Eq. (\ref{eq:Heff1}) we consider terms of the form
\bea
    \bra{k}\hat{H}_s^{eff}\ket{p} = \bra{k} e^{\frac{\lambda}{\Omega} (\hat{a}^{\dagger} - \hat{a})\hat{S}} \hat{H}_{s} e^{-\frac{\lambda}{\Omega} (\hat{a}^{\dagger} - \hat{a})\hat{S}} \ket{p}.
\eea
Note, $\ket{p} = \frac{1}{\sqrt{ p!}} (\hat{a}^{\dagger})^p \ket{0}$. 
We can again re-express the effective system Hamiltonian in terms of a ground state expectation value,
\bea
    \bra{k}\hat{H}_s^{eff}\ket{p} = \frac{1}{\sqrt{k! p!}} \bra{0} \hat{a}^k D(\alpha) \hat{H}_{s} D^{\dagger}(\alpha) (\hat{a}^{\dagger})^p \ket{0}.
\eea
Next we make use of the property of  displacement operators, $\ket{0} = D(\alpha)\ket{-\alpha}$. As such, we  rewrite our matrix element in terms of coherent state expectation values,
\bea
    \bra{k}\hat{H}_s^{eff}\ket{p} = \frac{1}{\sqrt{k! p!}} \bra{-\alpha} D^{\dagger}(\alpha) \hat{a}^k D(\alpha) \hat{H}_{s} D^{\dagger}(\alpha) (\hat{a}^{\dagger})^p D(\alpha) \ket{-\alpha}.
\eea
Furthermore, using yet another property of displacement operators: $D^{\dagger}(\alpha) \hat{a}^k D(\alpha) = (\hat{a} + \alpha)^k$, allows us to displace the RC operators,
\bea
    \bra{k}\hat{H}_s^{eff}\ket{p} = \frac{1}{\sqrt{k! p!}} \bra{-\alpha}  (\hat{a} + \alpha)^k  \hat{H}_{s} (\hat{a}^{\dagger} + \alpha^{\dagger})^p \ket{-\alpha}.
\eea
We note that the parameter $\alpha$ in our expressions is hermitian. Furthermore, we can express the coherent state in the basis of the harmonic oscillator number eigenstates $\ket{-\alpha} = \sum_{m=0}^{\infty} \frac{(-1)^m}{\sqrt{m!}} \alpha^m e^{-\frac{|\alpha|^2}{2}}$. 
Next, we use the binomial theorem to write $(\hat{a} + \alpha)^k = \sum_{l=0}^k \binom{k}{l} \hat{a}^l \alpha^{k-l}$. Combining these two manipulations, we write down the matrix element as 
%
\bea
    \bra{k}\hat{H}_s^{eff}\ket{p} = \frac{1}{\sqrt{k! p!}} \sum_{j=0}^p \sum_{l = 0}^k \sum_{n,m = 0}^{\infty} \frac{(-1)^{n+m}}{\sqrt{n! m!}} \binom{k}{l} \binom{k}{j} e^{-\frac{|\alpha|^2}{2}} \bra{n}  \hat{a}^l \alpha^{k-l+n} \hat{H}_{s} \alpha^{p-j+m}  (\hat{a}^{\dagger})^j\ket{m} e^{-\frac{|\alpha|^2}{2}}.
\eea
Note the action of the creation operator $(\hat{a}^{\dagger})^j\ket{m} = \sqrt{\frac{(m+j)!}{m!}} \ket{m+j}$. Therefore,
\bea
    \bra{k}\hat{H}_s^{eff}\ket{p} &=& \frac{1}{\sqrt{k! p!}} \sum_{j = 0}^p \sum_{l = 0}^k \sum_{n,m = 0}^{\infty} \frac{(-1)^{n+m} \sqrt{(m+j)!(n+l)!}}{{n! m!}} \binom{k}{l} \binom{p}{j} e^{-\frac{|\alpha|^2}{2}} \bra{n+l} \alpha^{k-l+n} \hat{H}_{s} \alpha^{p-j+m} \ket{m+j} e^{-\frac{|\alpha|^2}{2}}.
    \\
    &=& 
    \frac{1}{\sqrt{k! p!}} \sum_{j=0}^p \sum_{l = 0}^k \sum_{n,m = 0}^{\infty} \frac{(-1)^{n+m} \sqrt{(m+j)!(n+l)!}}{{n! m!}} \binom{k}{l} \binom{p}{j} e^{-\frac{|\alpha|^2}{2}} \alpha^{k-l+n} \hat{H}_{s} \alpha^{p-j+m} e^{-\frac{|\alpha|^2}{2}} \delta_{n+l,m+j}.
\eea
%
The kronecker-delta function implies that $m = n +l -j$. Re-indexing the sum over $m$, we arrive at our final expression
\bea
    \bra{k}\hat{H}_s^{eff}\ket{p} &=& \frac{1}{\sqrt{k! p!}} \sum_{j=0}^p \sum_{l = 0}^k \sum_{n=j-l}^{\infty} \frac{(-1)^{2n+l-j} (n+l)!}{{n! (n+l-j)!}} \binom{k}{l} \binom{p}{j} e^{-\frac{|\alpha|^2}{2}} \alpha^{k-l+n} \hat{H}_{s} \alpha^{p-2j+l+n} e^{-\frac{|\alpha|^2}{2}}
    \\
    &=&
    \frac{1}{\sqrt{k! p!}} \sum_{j = 0}^p \sum_{l = 0}^k \sum_{n=j-l}^{\infty} \frac{(-1)^{l-j} (n+l)!}{{n! (n+l-j)!}} \binom{k}{l} \binom{p}{j} \left(\frac{\lambda}{\Omega}\right)^{k+p-2j+2n} e^{-\frac{\lambda^2}{2\Omega^2}\hat{S}^2} \hat{S}^{k-l+n} \hat{H}_{s} \hat{S}^{p-2j+l+n} e^{-\frac{\lambda^2}{2\Omega^2}\hat{S}^2}.
    \label{eq:HeffK}
\eea
This expression can be readily computed to provide the higher order corrections to the effective system Hamiltonian.


\renewcommand{\theequation}{C\arabic{equation}}
\setcounter{equation}{0}  
\setcounter{section}{0} 
\section*{Appendix C: Equilibrium state of the generalized spin-boson: From asymptotically weak to ultra-strong}
\label{app:3}

In this appendix we provide further mathematical details on the effective mean force Gibbs state of the generalized spin-boson model discussed in Sec. \ref{sec:qubit}. In the main text, we expressed this state in the form of Eq. (\ref{eq:effMFGS}),
where $\hat{H}_s^{eff}(\lambda)$ is given in terms of a closed form expression Eq. (\ref{eq:HSBeff}). Using this Hamiltonian, we may then write the equilibrium state in a convenient form
\bea
e^{-\beta\hat{H}_s^{eff}(\lambda)} = e^{-\frac{1}{2}\beta \Delta (\Vec{v} \cdot \Vec{\sigma})},
\eea
where here $\vec{\sigma} = (\hat{\sigma}_x,\hat{\sigma}_y,\hat{\sigma}_z)$ and 
\bea
\nonumber
\vec{v} = [(1-e^{\frac{-2\lambda^2}{\Omega^2}})\sin(2\theta),0,(1+e^{\frac{-2\lambda^2}{\Omega^2}}) + (1-e^{\frac{-2\lambda^2}{\Omega^2}})\cos(2\theta)].
\\
\eea
Using properties of the Pauli operators, we may re-express the effective Gibbs state as
\bea
e^{-\beta\hat{H}_s^{eff}(\lambda)} = \cosh(\frac{\beta\Delta}{2} \vert \vec{v} \vert)\hat{I} - (\hat{v} \cdot \vec{\sigma})\sinh(\frac{\beta \Delta}{2} \vert \vec{v} \vert),
\eea
where, $\hat{v}$ is the unit vector associated to $\vec{v}$ and its magnitude is given by
\bea
\vert \vec{v} \vert = \sqrt{2(1+e^{-\frac{4\lambda^2}{\Omega^2}}) + 2(1-e^{-\frac{4\lambda^2}{\Omega^2}})\cos(2\theta)}.
\eea
Therefore, the partition function of the effective mean force Gibbs state is
\bea
Z_{eff} = \Tr{e^{-\beta \hat{H}_{s}^{eff}(\lambda)}} = 2\cosh(\frac{\beta\Delta}{2} \vert \vec{v} \vert).
\eea
As a result, we may write the equilibrium state of the system in a compact form as
\bea
    \rho_{eff}^{SS} = \frac{1}{2}\left[ \hat{I} - \frac{(\vec{v}\cdot\vec{\sigma})}{\vert \vec{v} \vert}\tanh(\frac{\beta\Delta}{2} \vert \vec{v} \vert)\right].
\eea
Writing explicitly the full $\lambda$ and $\theta$ dependence of this model, we obtain our final solution for the effective mean force Gibbs state of the generalized spin-boson model, valid for any coupling strength $\lambda$ from asymptotically weak to ultrastrong 
\bea
    \nonumber
    \rho_{eff}^{SS} = \frac{1}{2}\left[1 - \frac{(1-e^{\frac{-2\lambda^2}{\Omega^2}})\sin(2\theta)\hat{\sigma}_x + ((1+e^{\frac{-2\lambda^2}{\Omega^2}}) + (1-e^{\frac{-2\lambda^2}{\Omega^2}})\cos(2\theta))\hat{\sigma}_z }{\sqrt{2(1+e^{-\frac{4\lambda^2}{\Omega^2}}) + 2(1-e^{-\frac{4\lambda^2}{\Omega^2}})\cos(2\theta)}}\tanh(\frac{\beta\Delta}{2} \sqrt{2(1+e^{-\frac{4\lambda^2}{\Omega^2}}) + 2(1-e^{-\frac{4\lambda^2}{\Omega^2}})\cos(2\theta)})  \right].
    \label{eq:rhoeffLambda}
    \\
\eea
We highlight two limiting cases of Eq. (\ref{eq:rhoeffLambda}) where our results in Figure \ref{fig:Thermalization} and Figure \ref{fig:Ultrastrong} were validated. Namely: (i) the asymptotically weak coupling limit ($\lambda \to 0$), where we expect our solution to converge to a standard Gibbs state, and (ii) the ultrastrong coupling limit ($\lambda \to \infty$), where we expect our solution to agree with Eq. (\ref{eq:USlimit}). In these cases we obtain:
\bea
\lim_{\lambda \to 0} \rho_{eff}^{SS} = \frac{1}{2}(1 - \hat{\sigma}_z\tanh(\beta \Delta)) = e^{-\beta \Delta \sigma_z},
\eea
and
\bea
\lim_{\lambda \to \infty} \rho_{eff}^{SS} 
= \frac{1}{2}\left\{ 1 - \left[ \hat{\sigma}_x \sin(\theta) + \hat{\sigma}_z \cos(\theta) \right] \tanh(\beta\Delta\cos(\theta))\right\}.
\eea
Our asymptotically-weak coupling limit exactly agree with the standard Gibbs state.
Moreover, our ultrastrong result matches the ultrastrong limit of Ref. \citenum{Cresser}. Therefore, we prove analytically that the RCPT method generates effective Hamiltonian models that are exact in both the asymptotically weak and ultrastrong coupling regimes for the generalized spin-boson model. 

\renewcommand{\theequation}{D\arabic{equation}}
\setcounter{equation}{0}  
\setcounter{section}{0} 
\section*{Appendix D: Intermediate steps in the RCPT mapping of phonon-assisted electron transport}
\label{app:4}

 
 We start with the Hamiltonian (\ref{eq:Hephonon}) and describe its mapping to Eq. (\ref{eq:HephononEff}).
 First, we build from Eq. (\ref{eq:Hephonon}) the total RC Hamiltonian,
\bea
\hat{H}_{RC} &= &\epsilon_L \hat{L} + \epsilon_R \hat{R} + (\epsilon_L + \epsilon_R + U)\hat{D} + 
\Omega \left(\hat{a}^{\dagger} + \frac{\lambda}{\Omega}\hat{S} \right) 
\left(\hat{a} 
+ \frac{\lambda}{\Omega}\hat{S} \right) 
\nonumber\\
&+& \sum_q \omega_q \left( \hat{b}_q^{\dagger} + \frac{f_q}{\omega_q} (\hat{a}^{\dagger} + \hat{a}) \right) \left( \hat{b}_q + \frac{f_q}{\omega_q} (\hat{a}^{\dagger} + \hat{a}) \right) 
 \nonumber \\
&+& \sum_k \left[ \hat{A}_1 h_{k,L} \hat{c}_{k,L}^{\dagger} + \hat{A}_2 h^*_{k,L} \hat{c}_{k,L} \right ] + \sum_k \left [ \hat{A}_3 h_{k,R} \hat{c}_{k,R}^{\dagger} + \hat{A}_4 h^*_{k,R} \hat{c}_{k,R} \right] + \sum_k \epsilon_{k,L} \hat{c}_{k,L}^{\dagger} \hat{c}_{k,L} + \sum_k \epsilon_{k,R} \hat{c}_{k,R}^{\dagger} \hat{c}_{k,R}.
\label{eq:HephononRC}
\eea
In this expression, $\lambda$ is the coupling strength between the dots and the RC and $\Omega$ is the frequency of the RC.
$\hat{a}^{\dagger}$ ($ \hat{a}$) is the creation (annihilation) operator of the RC. 
The coupling energies between the RC and residual phononic bath modes of frequency $\omega_q$ 
are captured by $f_q$, while the 
creation (annihilation) operators of the residual phonon bath
are given by $\hat{b}^{\dagger}_q$ ($\hat{b}_q$).


Continuing with the RCPT procedure, we now apply a polaron transformation to partially decouple the phononic RC and the electronic dots, 
 $\hat{U}_P = e^{\frac{\lambda}{\Omega} \hat{S} (\hat{a}^{\dagger} - \hat{a})}$. 
This rotation results in the following transformed Hamiltonian, $\hat{H}_{RC-P} = \hat{U}_P \hat{H}_{RC} \hat{U}_P^{\dagger}$, given by
\bea
\nonumber
\hat{H}_{RC-P} = &&\epsilon_L \hat{U}_P\hat{L}\hat{U}_P^{\dagger} + \epsilon_R \hat{U}_P\hat{R}\hat{U}_P^{\dagger} + (\epsilon_L + \epsilon_R + U) \hat{U}_P\hat{D}\hat{U}_P^{\dagger} + \Omega \hat{a}^{\dagger}\hat{a} +  \sum_q \omega_q \left( \hat{b}_q^{\dagger} + \frac{f_q}{\omega_q}
 (\hat{a}^{\dagger} + \hat{a} - \frac{2\lambda}{\Omega}\hat{S})  \right) \left( \hat{b}_q + \frac{f_q}{\omega_q} (\hat{a}^{\dagger} + \hat{a} - \frac{2\lambda}{\Omega}\hat{S})  \right)
\\ \nonumber
&+& \sum_k \left[ \hat{U}_P\hat{A}_1\hat{U}_P^{\dagger} h_{k,L} \hat{c}_{k,L}^{\dagger} + \hat{U}_P\hat{A}_2\hat{U}_P^{\dagger} h^*_{k,L} \hat{c}_{k,L} \right ] + \sum_k \left [ \hat{U}_P\hat{A}_3\hat{U}_P^{\dagger} h_{k,R} \hat{c}_{k,R}^{\dagger} + \hat{U}_P\hat{A}_4\hat{U}_P^{\dagger} h^*_{k,R} \hat{c}_{k,R} \right] + \sum_k \epsilon_{k,L} \hat{c}_{k,L}^{\dagger} \hat{c}_{k,L} + \sum_k \epsilon_{k,R} \hat{c}_{k,R}^{\dagger} \hat{c}_{k,R}.
\nonumber\\
\label{eq:HephononRCPT}
\eea
Since the polaron transformation affects both the dots (through $\hat S$) and the RC (through $\hat a$),
terms affected by the polaron transformation  involve both the
RC Hilbert space and the double dot.
With regard to the RC, we make use of the fact that 
$\hat{U}_P \hat{a} \hat{U}_P^{\dagger} = \hat{a} - \frac{\lambda}{\Omega}\hat{S}$.
To compute terms that involve the quantum dots, 
we note that Eq. (\ref{eq:HSeff}) can be readily applied to any operator on the double dot Hilbert space. 
We give an example of the transformation to $\hat{L}$, noting that all other terms are computed in an analogous manner. Once the reaction coordinate is truncated to zero occupation, we obtain 
\bea
\nonumber
\bra{0} \hat{U}_P \hat{L} \hat{U}_P^{\dagger} \ket{0} = e^{-\frac{\lambda^2}{2\Omega^2} \hat{S}^2}\left( \sum_{n=0}^\infty \frac{\lambda^{2n}}{\Omega^{2n} n!} \hat{S}^n\hat{L} \hat{S}^n \right) e^{-\frac{\lambda^2}{2\Omega^2} \hat{S}^2}.
\nonumber\\
\eea
In this case, $\hat{S}^n \hat{L} \hat{S}^n$ is equal to $\hat{L}$ for $n$  even,
and to $\hat{R}$ for $n$ odd. Furthermore, $\hat{S}^2$ is diagonal in this case. 
An intermediate step in this derivation gives
\bea
\bra{0} \hat{U}_P \hat{L} \hat{U}_P^{\dagger} \ket{0} &=
& \cosh(\frac{\lambda^2}{\Omega^2}) e^{-\frac{\lambda^2}{2\Omega^2}\hat{S}^2} \hat{L} e^{-\frac{\lambda^2}{2\Omega^2}\hat{S}^2}
\nonumber\\
&+&  \sinh(\frac{\lambda^2}{\Omega^2}) e^{-\frac{\lambda^2}{2\Omega^2}\hat{S}^2} \hat{R} e^{-\frac{\lambda^2}{2\Omega^2}\hat{S}^2}.
\eea
Since $e^{-\frac{\lambda^2}{2\Omega^2}\hat{S}^2} = \ket{G}\bra{G} + e^{-\frac{\lambda^2}{2\Omega^2}}\ket{L}\bra{L} + e^{-\frac{\lambda^2}{2\Omega^2}}\ket{R}\bra{R} + \ket{D}\bra{D}$, 
the left dot gets altered with energy renormalization.
Furthermore, we obtain a new coupling between the left and right dots,
\bea
\nonumber
&&\bra{0} \hat{U}_p \hat{L} \hat{U}_p^{\dagger} \ket{0} =\cosh(\frac{\lambda^2}{\Omega^2})e^{-\frac{\lambda^2}{\Omega^2}}\hat{L} + \sinh(\frac{\lambda^2}{\Omega^2})e^{-\frac{\lambda^2}{\Omega^2}}\hat{R}.
\nonumber\\
\eea
%
For completeness, we  also show the transformation of $\hat{A}_1$, since it is distinct from the last computation. 
\bea
\nonumber
    \bra{0}\hat{U}_P \hat{A}_1 \hat{U}_P \ket{0} = e^{-\frac{\lambda^2}{2\Omega^2} \hat{S}^2}\left( \sum_{n=0}^\infty \frac{\lambda^{2n}}{\Omega^{2n} n!} \hat{S}^n\hat{A}_1 \hat{S}^n \right) e^{-\frac{\lambda^2}{2\Omega^2} \hat{S}^2}.
    \nonumber\\
\eea
Here, $\hat{S}^n \hat{A}_1 \hat{S}^n$ is equal to zero unless $n=0$. Therefore, 
\bea
    \bra{0}\hat{U}_P \hat{A}_1 \hat{U}_P \ket{0} = e^{-\frac{\lambda^2}{2\Omega^2}} \hat{A}_1.
\eea
%
%
The effective model is defined as
\bea
\hat{H}^{eff}(\lambda) = \bra{0}\hat{H}_{RC-P}\ket{0},
\eea
and we arrive at Eq. (\ref{eq:HephononEff}) in the main text.

\renewcommand{\theequation}{E\arabic{equation}}
\setcounter{equation}{0}  
\setcounter{section}{0} 
\section*{Appendix E: Intermediate steps in the RCPT mapping of Dissipative spin-chains}
\label{app:5}

In this Appendix, we begin from the model Hamiltonian 
(\ref{eq:Hspinchain}) and include the intermediate steps in deriving the effective model Hamiltonian Eq. (\ref{eq:H2Qeff}).

Starting with Eq. (\ref{eq:Hspinchain}),
we apply the RCPT method and extract a reaction coordinate from each reservoir,
%
\bea
    \hat{H}_{RC} &= &\sum_{\alpha=1}^N \Delta_{\alpha} \hat{\sigma}_z^{\alpha} 
    + \sum_{i\in\{x,y,z\}} \sum_{\alpha=1}^{N-1}J_i \hat{\sigma}_i^{\alpha} \hat{\sigma}_i^{\alpha+1} + \sum_{\alpha=1}^{N} \Omega_{\alpha} \left(\hat{a}_{\alpha}^{\dagger} + \frac{\lambda_{\alpha}}{\Omega_{\alpha}} \hat{\sigma}_x^{\alpha} \right) \left(\hat{a}_{\alpha}^{\dagger} + \frac{\lambda_{\alpha}}{\Omega_{\alpha}} \hat{\sigma}_x^{\alpha} \right) 
     \nonumber\\
    &+& \sum_{\alpha=1}^N\sum_k \omega_{\alpha,k} \left( \hat{b}_{\alpha,k}^{\dagger} + \frac{f_{\alpha,k}}{\omega_{\alpha,k}}(\hat{a}_{\alpha}^{\dagger} + \hat{a}_{\alpha})\right) \left( \hat{b}_{\alpha,k} + \frac{f_{\alpha,k}}{\omega_{\alpha,k}} (\hat{a}_{\alpha}^{\dagger} + \hat{a}_{\alpha}) \right),
\eea
%
with $\lambda_{\alpha}$ and $\Omega_{\alpha}$ denoting the coupling strength and frequency of the $\alpha$th reaction coordinate.
Next, we perform a polaron transformation on each RC, since the unitary operators commute. Explicitly, it is given as
\bea
    \hat{U}_P = \Pi_{\alpha=1}^{N}\hat{U}_{P,\alpha} 
    = \Pi_{\alpha=1}^{N} e^{\frac{\lambda_{\alpha}}{\Omega_{\alpha}}(\hat{a}_{\alpha}^{\dagger}-\hat{a}_{\alpha})\hat{\sigma}_x^{\alpha}}.
\eea
We pause here to note that studying multi-qubit systems is natural for the RCPT because the qubits operate on different Hilbert spaces, and $\hat{S}^2 = 1$. As a result, performing multiple polaron transformations and generating the effective model is relatively simple compared to the other models studied in this work. We apply the polaron transformation and arrive at the following Hamiltonian,
%
\bea
    \hat{H}_{RC-P} &= &\sum_{\alpha=1}^{N} \Delta_{\alpha}  \hat{U}_P \hat{\sigma}_z^{\alpha} \hat{U}_P^{\dagger} 
    + \sum_{i\in\{x,y,z\}}\sum_{\alpha=1}^{N-1} J_i \hat{U}_P \hat{\sigma}_i^{\alpha} \hat{\sigma}_i^{\alpha+1} \hat{U}_P^{\dagger} + \sum_{\alpha=1}^{N} \Omega_{\alpha} \hat{a}_{\alpha}^{\dagger}\hat{a}_{\alpha}
     \nonumber\\
    &+& \sum_{\alpha=1}^N\sum_k \omega_{{\alpha},k} \left( \hat{b}_{{\alpha},k}^{\dagger} + \frac{f_{\alpha,k}}{\omega_{\alpha,k}} (\hat{a}_{\alpha}^{\dagger} + \hat{a}_{\alpha} - \frac{2\lambda_{\alpha}}{\Omega_{\alpha}} \hat{\sigma}_x^{\alpha})\right) \left( \hat{b}_{\alpha,k} + \frac{f_{\alpha,k}}{\omega_{\alpha,k}} (\hat{a}_{\alpha}^{\dagger} + \hat{a}_{\alpha} - \frac{2\lambda_{\alpha}}{\Omega_{\alpha}} \hat{\sigma}_x^{\alpha}) \right).
\eea
%
Next, we introduce a shorthand notation where the ket vector $\ket{\bf{0}} = \ket{0_1,0_2,...,0_{N}}$
denotes a zero excitation state of each RC. Since the polaron transformations act on different Hilbert spaces, we can simply apply each truncation separately. 
%
Therefore, we should only evaluate the action of the polaron transformation on each Pauli operator $\hat{\sigma}_{x,y,z}$, done by employing Eq. (\ref{eq:HSeff}).
First, we note that $[\hat{U}_P ,\hat{\sigma}_x^{\alpha}] = 0$, so the action of the polaron transformation is trivial on $\hat{\sigma}_x$. 
The transformations of $\hat{\sigma}_z$ and $\hat{\sigma}_y$ are very similar, and produce the same outcome. We demonstrate the action on $\hat{\sigma}_y$, noting that we computed analogous expressions for $\hat{\sigma}_z$ in Sec. \ref{sec:qubit},
\bea
&&\bra{\bf{0}} \hat{U}_{P,\alpha} \hat{\sigma}^{\alpha}_y \hat{U}_{P,\alpha}^{\dagger} \ket{\bf{0}} =  e^{-\frac{\lambda_{\alpha}^2}{\Omega_{\alpha}^2}} \sum_n \frac{\lambda^{2n}_{\alpha}}{\Omega^{2n}_{\alpha} n!} (\hat{\sigma}_x^{\alpha})^n \hat{\sigma}_y^{\alpha} (\hat{\sigma}_x^{\alpha})^n 
 \nonumber\\
&&= e^{-\frac{\lambda_{\alpha}^2}{\Omega_{\alpha}^2}} \sum_{n,even} \frac{\lambda^{2n}_{\alpha}}{\Omega^{2n}_{\alpha} n!} \hat{\sigma}_y^{\alpha} + \sum_{n, odd} \frac{\lambda^{2n}_{\alpha}}{\Omega^{2n}_{\alpha} n!} \hat{\sigma}_x^{\alpha} \hat{\sigma}_y^{\alpha} \hat{\sigma}_x^{\alpha}
 \nonumber\\
&&= e^{-\frac{\lambda_{\alpha}^2}{\Omega_{\alpha}^2}} \left[\cosh(\frac{\lambda_{\alpha}^2}{\Omega_L^2}) - \sinh(\frac{\lambda_{\alpha}^2}{\Omega_{\alpha}^2})\right] \hat{\sigma}_y^{\alpha}
 \nonumber\\
&&= e^{-\frac{2\lambda_{\alpha}^2}{\Omega_{\alpha}^2}} \hat{\sigma}_y^{\alpha}.
\eea
%
The effect of strong system-bath coupling as seen from the RCPT method in this model is again parameter renormalization in both the spin splittings as well as the internal interactions:
$\Delta_{\alpha} \xrightarrow[]{} \Delta_{\alpha} e^{-\frac{2\lambda_{\alpha}^2}{\Omega_{\alpha}^2}}$,
$J_x \xrightarrow[]{} J_x$,
$J_y \xrightarrow[]{} J_y e^{-\frac{2\lambda_{\alpha}^2}{\Omega_{\alpha}^2}} e^{-\frac{2\lambda_{\alpha+1}^2}{\Omega_{\alpha+1}^2}}$,
$J_z \xrightarrow[]{} J_z e^{-\frac{2\lambda_{\alpha}^2}{\Omega_{\alpha}^2}} e^{-\frac{2\lambda_{\alpha+1}^2}{\Omega_{\alpha+1}^2}}$.
%
The effective model is defined as
\bea
\hat{H}^{eff}(\lambda_1,\lambda_2,....\lambda_N) = \bra{0_1,0_2,...0_N}\hat{H}_{RC-P}\ket{0_1,0_2,....,0_N},
\eea
and we arrive at Eq. (\ref{eq:H2Qeff}) in the main text.

\end{widetext}


\end{document}